\documentclass[]{article}   

\usepackage{bm}
\usepackage[spanish,english]{babel}
\usepackage{amsfonts}
\usepackage{amssymb}
\usepackage{graphicx}
\usepackage[latin1]{inputenc}
\usepackage{hyperref}

\newcommand{\LL}{\mathcal{L}}

\newcommand{\be}{\begin{equation}}
\newcommand{\ee}{\end{equation}}
\newcommand{\bea}{\begin{eqnarray}}
\newcommand{\eea}{\end{eqnarray}}

\begin{document}

\title{Born-Infeld-$f(R)$ gravity}   

\author{Andrey N. Makarenko$^{1,2}$,  Sergei Odintsov$^{3,4,5}$, Gonzalo J. Olmo$^{6,7}$ \\  \\ {\small $^1$Tomsk State Pedagogical University, ul. Kievskaya, 60, 634061 Tomsk, Russia} \\ {\small 
$^2$National Research Tomsk State University, Lenin Avenue, 36, 634050 Tomsk, Russia} \\  
{\small $^3$Instituci\`{o} Catalana de Recerca i Estudis Avancats (ICREA), Barcelona, Spain} \\ 
{\small  
$^4$Institut de Ciencies de l'Espai (CSIC-IEEC), Campus UAB,}\\{\small 
 Torre C5-Par-2a- pl, E-08193 Bellaterra (Barcelona), Spain} \\  
{\small $^5$ King Abdulaziz University, Jeddah, Saudi Arabia}\\
{\small $^6$Depto. de F\'{i}sica Te\'{o}rica \& IFIC , Universidad de Valencia - CSIC} \\ {\small  Burjassot 46100, Valencia, Spain }\\ 
{\small $^7$Depto. de F\'isica, Universidade Federal da
Para\'\i ba, 58051-900 Jo\~ao Pessoa, Para\'\i ba, Brazil}}

\date{March, 2014}    

\maketitle

\begin{abstract}
Motivated by the properties of matter quantum fields in curved space-times, we work out a gravity theory that combines the Born-Infeld gravity Lagrangian with an $f(R)$ piece. To avoid ghost-like instabilities, the theory is formulated within the Palatini approach. This construction provides more freedom to address a number of important questions such as the dynamics of the early universe and the cosmic accelerated expansion, among others. In particular, we consider the effect that adding an $f(R)=a R^2$  term has on the early-time cosmology. We find that bouncing solutions are robust against these modifications of the Lagrangian whereas the  solutions with {\it loitering} behavior of the original Born-Infeld theory are very sensitive to the $R^2$ term. In fact, these solutions are modified in such a way that a plateau in the $H^2$ function may arise yielding a period of (approximately) de Sitter inflationary expansion. This inflationary behavior may be found even in a radiation dominated universe.  
\end{abstract}

\section{Introduction}

Extensions of General Relativity (GR) have been considered in the literature following different approaches and motivated by a variety of reasons. Theoretical arguments support that GR is just an effective theory that fits well the behavior of gravitational systems at relatively low energies. At ultrahigh and at very low energies or, equivalently, at ultrashort and very large length scales, corrections to the GR Lagrangian are expected. The form of these corrections is difficult to guess from first principles and probably results from complicated processes related to the fundamental constituents and/or structure of space-time and how their symmetries are broken. Moreover, there is no experimental evidence whatsoever about what is the most reasonable or favourable formulation of classical GR that should be used to consider its high-energy and low-energy extensions. What should be the classical starting point? Should we stick to the traditional metric (or Riemannian) approach or should we consider a Palatini (or metric-affine) formulation?  Whatever the choice, the potential extensions offered by each starting point can lead to significantly different gravitational physics. 

In this sense, it is well-known that high-curvature extensions of GR in the usual metric formalism generically lead to higher-order derivative equations and/or to the emergence of new dynamical degrees of freedom. This is the case, for instance, of $f(R)$ theories \cite{review0,review1,review2,review3,review4,Sotiriou:2008rp}, quadratic gravity, and the Born-Infeld type gravity action considered by Deser and Gibbons \cite{Deser:1998rj}, to name just a few. If a Palatini formulation of those theories is chosen, however, one finds completely different physics \cite{Banados}. In fact, it is well established that in the Palatini approach those theories lead to second-order  metric field equations which in vacuum exactly recover the dynamics of GR \cite{olmo11}. 
 
It turns out that, in the above mentioned Palatini theories, despite the fact of allowing the connection to vary independently of the metric, the number of dynamical fields ends up being the same as in standard GR. One finds that the connection can be solved in terms of the metric and the matter sources via a set of algebraic (not differential) equations. Leaving aside the dependence on the matter, this is exactly what happens in the Palatini formulation of GR, where the connection becomes a constrained object algebraically related with the first derivatives of the metric, thus defining the Levi-Civita connection. Therefore, even though one might {\it a priori} expect many new additional degrees of freedom in the metric-affine formulation due to the independence of the connection, the resulting equations yield a different answer, namely, that the connection is not a dynamical object and that the metric satisfies second-order equations.  Moreover, in general, one finds that in the absence of matter fields the metric field equations exactly boil down to those of GR with an effective cosmological constant (see \cite{olmo11} for further details and discussions). The modified dynamics of these theories, therefore, is not generated by new dynamical degrees of freedom, which has motivated recent related research on theories with non-dynamical fields \cite{Bazeia:2014rea,Pani:2013qfa}. A closer inspection of the Palatini models puts forward that the modified dynamics is due to nonlinearities induced by the matter sources and by higher-order spatial derivatives of the fields \cite{Olmo:2011fh,Harko:2014zma}.  \\

The fact of having second-order equations so closely related to GR is of great importance \cite{Fiorini} because it minimizes the number of extra inputs necessary to characterize a given solution 
simply because higher-order equations require more boundary/initial conditions.  
In the Palatini version of the Born-Infeld gravity model, for instance, there is no more freedom than in GR to get rid of cosmic singularities starting from a solution which asymptotes to our current accelerated expansion phase. If the big bang singularity is avoided in this model, it is because the theory is doing something robust and relevant on the dynamics, not because we have extra freedom to select a subset of solutions in an {\it ad hoc} manner, 
as it happens in theories with higher-order derivatives.  
This type of theories, therefore, must be explored in more detail, as the modified dynamics they generate is enough to successfully avoid important problems without any further external or {\it ad hoc} input. Quadratic gravity is also able to avoid the big bang singularity \cite{Barragan:2010qb,Barragan:2009sq}.  
 When the Palatini version is considered, this occurs in a purely dynamical way with exactly the same number of initial conditions (at late times) as in GR. In the metric version of the theory \cite{Anderson, Novello}, however, additional restrictions on the parameters that characterize the asymptotically FRW solutions are necessary.  

The Born-Infeld (BI) gravity model is a very interesting starting point to consider high-energy extensions of GR because BI-like Lagrangians naturally arise in different scenarios in a very fundamental way. For instance, the original Born-Infeld theory replaced the classical Maxwell Lagrangian $L_M=-\frac{1}{16\pi}F_{\mu\nu}F^{\mu\nu}$ by a new version 
\begin{equation}
\LL_{BI}= \frac{\beta^2}{8\pi}\left(\sqrt{-\vert \eta_{\mu\nu} + \beta^{-1} F_{\mu\nu} \vert} - 1 \right) \label{eq:BIem0} \ ,
\end{equation}
which for a pure electric field can also be written as 
\begin{equation}
\LL_{BI}= \frac{\beta^2}{8\pi}\left(\sqrt{1+\frac{F_{\mu\nu}F^{\mu\nu}}{\beta^2}} - 1 \right) \label{eq:BIem1} \ . 
\end{equation}
This new theory sets an upper bound for the electric field strength and regularizes the energy of a point particle, which is divergent in the standard Maxwell theory. On the other hand, the modification needed in the Lagrangian of a free point-particle to go from a non-relativistic, $L_{nr}=\frac{1}{2}mv^2$, to a relativistic description is also of the BI type \cite{Ferraro:2009zk}: $L_{rel}=mc^2(1-\sqrt{1-\frac{mv^2}{mc^2}})$. In analogy with (\ref{eq:BIem0}), Deser and Gibbons \cite{Deser:1998rj} proposed a Born-Infeld like theory of gravity which has been recently reconsidered by Ba\~{n}ados and Ferreira \cite{Banados} in the Palatini formulation,
\begin{equation}\label{eq:A0}
S_{BI}=\frac{1}{\kappa^2\epsilon}\int d^4x \left[\sqrt{-|g_{\mu\nu}+\epsilon R_{\mu\nu}(\Gamma)|}-\lambda \sqrt{-|g_{\mu\nu}|}\right]+S_m \ ,
\end{equation}
as it yields second-order equations and avoids ghost-like instabilities. Here $g_{\mu\nu}$ represents the (non-flat) space-time metric and $R_{\mu\nu}(\Gamma)$ the Ricci tensor of the independent connection (further notational details later).

It is worth noting that  the Born-Infeld electromagnetic Lagrangian is consistent with the one-loop version of supersymmetric QED \cite{susyQED}. Additionally, the Lagrangians describing the electromagnetic field of certain D-branes are also of the Born-Infeld determinantal type \cite{Gibbons:2001gy}.  Therefore, this type of Lagrangians appear in a very fundamental way in different scenarios of interest. The possibility of regularizing curvature scalars in gravity via this type of Lagrangians has motivated a burst of activity in the context of Born-Infeld gravity in cosmological scenarios, where  the growth of perturbations, the effects on the angular power spectrum of the cosmic microwave background, and other aspects of scalar and tensorial linear perturbations and inflation have been investigated \cite{Du:2014jka, Kim:2013noa,Kruglov:2013qaa, Yang:2013hsa,Avelino:2012ue, DeFelice:2012hq, EscamillaRivera:2012vz, Cho:2012vg, Scargill:2012kg, EscamillaRivera:2013hv}. Other relevant questions dealing with astrophysics \cite{Harko:2013xma,Avelino:2012ge}, stellar structure \cite{Sham:2013cya,Kim:2013nna,Harko:2013wka,Sham:2013sya, Avelino:2012qe, Sham:2012qi,Pani:2012qd,  Pani:2011mg}, the problem of cosmic singularities \cite{Bouhmadi-Lopez:2013lha, Ferraro:2010at}, black holes \cite{Olmo:2013gqa}, and wormhole physics \cite{Lobo:2014fma,Harko:2013aya} have also been considered in the literature. 

 From an observational perspective, we note that BI theory recovers GR with an effective cosmological constant at the zeroth order in a series expansion in the parameter $\epsilon$. For this reason the theory can be made to agree with all current observations by just suitably tuning this parameter. Since we are mainly interested in theoretical aspects concerning the avoidance of singularities and alternative mechanisms for inflation, we will always assume that $\epsilon$ is sufficiently small so as not to enter in conflict with observations. \\

Despite the appealing properties of the BI gravity Lagrangian, our ignorance on the behavior of gravity at the highest energies motivates the exploration of departures from that basic structure to check the robustness of its predictions. In fact, if quantum effects in curved space-time are considered \cite{Parker-Toms}, in general one finds curvature corrections that are necessary to account for the renormalizability of matter fields in such backgrounds. These corrections are known to be quadratic in the Ricci and Riemann tensors at high energies, but other types of $R-$dependent corrections may arise in the infrared, thus having a relevant impact on the late-time cosmic expansion \cite{PARKER1,PARKER2}. This fact has also motivated recent studies of hybrid scenarios in which the Einstein-Hilbert Lagrangian is supplemented with $f(R)$ corrections of the Palatini type \cite{Harko:2011nh,Capozziello:2012ny,Capozziello:2012hr,Capozziello:2012qt,Capozziello:2013uya, Capozziello:2013yha,Capozziello:2013gza}.

The quantum properties of matter fields in curved space-times, therefore, naturally justify the interest in exploring high-energy and low-energy modifications of the classical BI theory via $f(R)$-type terms. In this sense, as advanced above, the BI gravity Lagrangian yields a low-energy perturbative expansion with GR as the lowest order followed by quadratic and higher-order curvature corrections with specific coefficients, which is in consonance with the expected quantum field theory corrections at high energies. Theories of this type, with up to quadratic curvature corrections, have been investigated within the Palatini approach in the literature \cite{or12a,or13d,PLB13,lor,Olmo:2013mla,Lobo:2014zla,Lobo:2014fma}, and specific methods to deal with the resulting field equations have been developed \cite{OSAT}. However,  higher-order curvature corrections involving cubic powers or higher of the Ricci tensor (such as $R_{\mu\alpha}R_{\beta\gamma}R_{\delta\nu}g^{\alpha\beta}g^{\gamma\delta}g^{\mu\nu}$, for instance) have not been explored yet and are likely to require new methods.   
By contrast, though the BI theory contains terms of that kind in a perturbative expansion,  in its exact determinantal form the methods required to deal with the field equations are much simpler even than for the quadratic theory.  
It is thus far from clear how a theory with a similar perturbative expansion as the BI theory but with different coefficients multiplying the higher-order curvature terms or including low-curvature corrections, like in the case of $f(R)$ theories,  could be put in a form amenable to calculations. In other words, slight modifications of the action possibly generated by the quantum properties of the matter fields can lead to non-trivial changes in the structure of the field equations, which may substantially difficult the analysis. In particular, if an $f(R)$ piece is added to the BI action (\ref{eq:A0}), one finds that the connection equation cannot be solved by just using the tensor $q_{\mu\nu}=g_{\mu\nu}+\epsilon R_{\mu\nu}$ as an auxiliary metric, and more elaborate manipulations are necessary in general. In this work we consider this problem in detail and extend the existing methods to deal with $f(R)$-like modifications of the field equations. This will allow us to explore, in particular, if the high-energy behavior of the BI theory itself is robust against small changes in the coefficients that define its perturbative series expansion. Recall, in this sense, that in curved space-times \cite{Parker-Toms,Anderson} the coefficients of the high-curvature corrections depend on the number  and spin of the matter fields.  \\

Taking a cosmological scenario with perfect fluids, we provide an algorithm that allows to efficiently study $f(R)$ departures from the original BI gravity theory in a fully non-perturbative way. This aspect, namely, the exact (non-perturbative) treatment of the field equations, is very important because the field equations of Palatini theories usually involve algebraic relations which must be handled with care in order not to miss important physical information (see, for instance, the discussion in the introduction of \cite{Ashtekar} regarding the properties of nonperturbative systems).  In fact, the replacement of the big bang singularity by a cosmic bounce and of black hole singularities by wormholes \cite{Lobo:2014fma,Olmo:2013gqa} in Palatini theories are non-perturbative properties that need not respond linearly to small modifications in the parameters of the theory. 

With the technical aspects of these BI-$f(R)$ theories under control, as an illustration, we study the robustness of the nonsingular cosmic solutions against modifications of the quadratic curvature terms. We confirm that the bouncing solutions of the original BI theory persist even for large variations in the coefficients of the perturbative expansion and find that the other kind of non-singular solutions, which represent a minimum volume in unstable equilibrium, may develop a big bang singularity followed by a period of approximately de Sitter expansion due to a plateau in the Hubble function. Unstable equilibrium configurations also arise for certain values of the equation of state. \\

The content is organized as follows. In section \ref{sec:field_eqs} we introduce the BI-$f(R)$ theory, derive the field equations, and put them in a form amenable to calculations. In section 
\ref{sec:fluids} we discuss the procedure to deal with perfect fluids, which will be used in a cosmological scenario in section  \ref{sec:cosmology}, where the main physical results are obtained. We conclude in section \ref{sec:summary} with a summary of the work and a discussion of the results.

\section{Born-Infeld-$f(R)$ gravity in Palatini formalism. \label{sec:field_eqs}}
Let us consider the following action made out of the Born-Infeld (BI) theory plus an $f(R)$ term
\begin{equation}\label{eq:A1}
S_{BI}=\frac{1}{\kappa^2\epsilon}\int d^4x \left[\sqrt{-|g_{\mu\nu}+\epsilon R_{\mu\nu}(\Gamma)|}-\lambda \sqrt{-|g_{\mu\nu}|}\right]+\frac{\alpha}{2\kappa^2}\int d^4x \sqrt{-g}f(R)+S_m \ ,
\end{equation}
where $g_{\mu\nu}$ represents the space-time metric, $R_{\mu\nu}(\Gamma)$ is the Ricci tensor of the connection $\Gamma^\alpha_{\beta\gamma}$, which is {\it a priori} independent of the metric (Palatini formalism), $\lambda$ is a constant or order unity, $f(R)$ is an unspecified function of the Ricci scalar $R=g^{\mu\nu}R_{\nu\mu}(\Gamma)$, and $S_m$ represents the matter action, which is only coupled to the metric as dictated by the equivalence principle. 

In the limit $\epsilon \to 0$, the BI Lagrangian recovers the usual GR term and the above action boils down to an $f(R)$ theory with Lagrangian $\LL_G=\frac{R-2\Lambda+\alpha f(R)}{2\kappa^2}$, where $\Lambda\equiv (\lambda-1)/\epsilon$. If instead we take the limit $\alpha\to 0$, we recover the BI theory. When $\alpha\to 0$ and $\epsilon\to 0$ GR is naturally recovered. In this action we assume vanishing torsion and a symmetric Ricci tensor. \\

The field equations follow from (\ref{eq:A1}) by independent variation with respect to the metric and the connection (Palatini formalism). The metric variation yields 
\begin{equation}\label{eq:g-var0}
\frac{\sqrt{-q}}{\sqrt{-g}}q^{\mu\nu}-\left[\left(\lambda-\frac{\alpha\epsilon }{2}f\right)g^{\mu\nu}+\alpha\epsilon f_R g^{\mu\beta} g^{\nu\gamma}R_{\beta\gamma}\right]=-\kappa^2\epsilon T^{\mu\nu} \ ,
\end{equation}
whereas the connection variation boils down to 
\begin{equation}\label{eq:con-var0}
\nabla_\beta\left[\sqrt{-q}q^{\mu\nu}+\alpha f_R\sqrt{-g}g^{\mu\nu}\right]=0 \ ,
\end{equation}
where the covariant derivative is defined in terms of the independent connection $\Gamma^\alpha_{\beta\gamma}$. By solving this equation one obtains the explicit form of  $\Gamma^\alpha_{\beta\gamma}$, which in general differs from the usual Christoffel symbols of the metric $g_{\mu\nu}$. 
In the above equations, we have used the notation
\begin{equation}\label{eq:q}
q_{\mu\nu}=g_{\mu\nu}+\epsilon R_{\mu\nu}(\Gamma) \ .
\end{equation}
The inverse of $q_{\mu\nu}$ has been denoted $q^{\mu\nu}$, and its form will be obtained explicitly later. The procedure to obtain $q^{\mu\nu}$ in a way consistent with the field equations is complicated and deserves a bit of previous discussion. 

\subsection{The conformal approach.}

It is well-known in the literature of Palatini $f(R)$ theories that the independent connection of the theory can be solved in terms of an auxiliary metric $h_{\mu\nu}$ which is conformal with $g_{\mu\nu}$ (for details see the review \cite{olmo11}). One can thus be tempted to proceed in a similar way with the BI-$f(R)$ theory presented here. We will see that such an approach is incomplete and leads to strong limitations. This indicates that a more general scenario must be considered, which is worked out in detail in Sec.\ref{sec:consistent}. Nonetheless, we include here a brief discussion of this point to illustrate its implications. 

Assume for now that $q_{\mu\nu}=p(R)g_{\mu\nu}$, with $p(R)$ a function of the Ricci scalar, and insert this ansatz into (\ref{eq:con-var0}), which yields
\begin{equation}
\nabla_\beta\left[(p(R)+\alpha f_R)\sqrt{-g}g^{\mu\nu}\right]=0 \ .
\end{equation}
We can now define an auxiliary tensor $u_{\mu\nu}=(p(R)+\alpha f_R)g_{\mu\nu}$ such that the above equation boils down to $\nabla_\beta\left[\sqrt{-u}u^{\mu\nu}\right]=0$. In Einstein's theory the connection equation takes exactly this form, $\nabla_\beta\left[\sqrt{-g}g^{\mu\nu}\right]=0$, which establishes the compatibility of the connection with the metric thus leading to the Levi-Civita connection as a solution (see \cite{MTW} for details) in the torsionless case. Therefore, in our case we have
\begin{equation}
\Gamma^\alpha_{\mu\nu}=\frac{1}{2} u^{\alpha\beta}\left(\partial_\mu u_{\nu\beta}+\partial_\nu u_{\mu\beta}-\partial_\beta u_{\mu\nu}\right).
\end{equation}
This provides a complete and exact solution of the connection equation. There remains, however, to determine the form of the function $p(R)$ and verify if this ansatz is valid for arbitrary $f(R)$, which requires the use of the other field equations. Now, confronting the conformal ansatz  with the definition (\ref{eq:q}), it follows that we are restricting ourselves to those cases in which $ R_{\mu\nu}(\Gamma)$ is proportional to $g_{\mu\nu}$. To be precise, one finds that $ \epsilon R_{\mu\nu}(\Gamma)=(p(R)-1)g_{\mu\nu}$. In a cosmological scenario, with line element $ds^2=-dt^2+a^2(t)\delta_{ij}dx^i dx^j$, one can verify that this relation imposes tight constraints on both functions $p(R)$ and $f(R)$. To see this, let us denote $u(t)\equiv (p(R)+\alpha f_R)$ and $r(t)=(p(R)-1)/\epsilon$. One then finds that $R(u_{\alpha\beta})_{\mu\nu}=r(t)g_{\mu\nu}$ leads to (overdots denote time derivative)
\begin{eqnarray}
r(t)&=&\frac{3}{2} \left[2\frac{\ddot{a}}{a}+\frac{\dot{a}}{a}\frac{\dot{u}}{u}+\frac{\ddot{u}}{u}-\left(\frac{\dot{u}}{u}\right)^2\right] \\
r(t)&=&  \left[\frac{\ddot{a}}{a}+\frac{5}{2}\frac{\dot{a}}{a}\frac{\dot{u}}{u}+\frac{\ddot{u}}{2u}+2\left(\frac{\dot{a}}{a}\right)^2\right]  \ ,
\end{eqnarray}
which can be combined to get ($H\equiv \dot a /a$)
\begin{eqnarray}\label{eq:Hubble-conf}
r(t)&=& 3\left(H+\frac{\dot{u}}{2u}\right)^2 \\ 
2\dot H&=& H \frac{\dot{u}}{u}+\frac{3}{2}\left(\frac{\dot{u}}{u}\right)^2-\frac{\ddot{u}}{u} \ .
\end{eqnarray}
Using these two equations, one can verify that $\frac{\dot{u}}{u}=\frac{\dot{r}}{r}$, which leads to 
\begin{equation}\label{eq:r=u}
r(t)=C u(t) \ ,
\end{equation}
with $C$ a constant. Now, combining the conformal ansatz $q_{\mu\nu}=p(R)g_{\mu\nu}$  with (\ref{eq:q}), one obtains that $p(R)=1+\epsilon R/4$.  Inserting this form of $p(R)$ into (\ref{eq:r=u}), one finds that the function $f(R)$ must take the form
\begin{equation}\label{eq:f(R)-conf}
f(R)=\frac{ (1-C \epsilon )}{8 \alpha  C}R^2 -\frac{R}{\alpha }+\Lambda \ ,
\end{equation}
where $\Lambda$ is an integration constant. We thus see that the conformal ansatz selects specific forms of the functions $p(R)$ and $f(R)$ and is, therefore, of limited interest. \\
On the other hand,  the conformal ansatz  together with  
 (\ref{eq:g-var0}) implies that 
\begin{equation}\label{eq:g-var0}
p(R) g^{\mu\nu}-\left[\left(\lambda-\frac{\alpha\epsilon }{2}f\right)g^{\mu\nu}+\alpha f_R (p(R)-1)g^{\mu\nu}\right]=-\kappa^2\epsilon T^{\mu\nu} \ .
\end{equation}
Substituting the form of the function  $f(R)$ obtained in (\ref{eq:f(R)-conf}), one gets  
and energy-momentum tensor in the form of a perfect fluid with energy density $\rho=\frac{2-2\lambda+\alpha \epsilon \Lambda}{2\kappa^2\epsilon}$ and pressure $P=-\rho$. One can verify that the vacuum case corresponds to $\Lambda=\frac{2(\lambda-1)}{\alpha\epsilon}$.

On the other hand, from the above expression (\ref{eq:Hubble-conf}), one obtains an equation relating  $H$ and $u$ as
\begin{equation}
\label{eeq1}
H=\pm\sqrt{\frac{u\,C}{3}}-\frac{\dot{u}}{2u}.
\ee
Note that the constant $C$ could be absorbed into a redefinition of the time coordinate. 
For constant $H=h$, one has $a=e^{ht}$, which leads to
\be
u= \frac{9 \left(3 e^{6 h c_1} h^2+c e^{2 h t+3 h c_1} h^2\pm 2 \sqrt{3} c e^{\frac{1}{2} (2 h t+9 h c_1)} h^2\right)}
{C(e^{4 h t}+9 e^{6 h c_1}-6 e^{2 h t+3 hc_1})}
\ee
One would thus conclude that the metric associated with the Christoffel symbols would be defined by the expression
\be
u_{\mu\nu}=
\frac{9 \left(3 e^{6 h c_1} h^2+c e^{2 h t+3 h c_1} h^2\pm 2 \sqrt{3} c e^{\frac{1}{2} (2 h t+9 h c_1)} h^2\right)}{C(e^{4 h t}+9 e^{6 h c_1}-6 e^{2 h t+3 hc_1})}g_{\mu\nu},
\ee
where $g_{\mu\nu}=$diag$(-1,\,\,e^{2ht}, \,\,e^{2ht},\,\,e^{2ht})$. Had one defined instead $u=e^{h\,t}$ (or $u=u_0 t^h$), then from the equation (\ref{eeq1}) the scale factor would be $a= a_0 e^{\frac{2  \sqrt{c} e^{\frac{h t}{2}}}{\sqrt{3} h}-\frac{h t}{2}}$ (or $a=a_0 e^{\frac{2 t \sqrt{\frac{u_0 t^h}{c}}}{\sqrt{3} (2+h)}} t^{-h/2}$). 
Thus, by specifying one of the metrics,  the other is automatically determined without explicit knowledge of the matter sources, which puts forward the peculiar properties of the conformal ansatz. In the appendices  \ref{App0} and  \ref{App1} it is shown that a conformal ansatz in a different theory and a non-conformal ansatz for (\ref{eq:A1}) can also constrain the form of the $f(R)$ function.

\subsection{Consistent manipulation of the field equations. \label{sec:consistent}}

We have just seen that imposing a conformal ansatz, which is the natural procedure in the case of pure $f(R)$ theories, leads to undesired restrictions on the family of theories one would like to consider. Now we show that the connection equation can be solved in a way that does not  impose any constraint on  the form of the function $f(R)$ that defines the gravity Lagrangian. This approach is fully consistent with the set of  metric and connection field equations and requires going beyond the conformal relation between metrics. \\

Using the notation $\hat q$ and $\hat q^{-1}$ to denote $q_{\mu\nu}$ and  $q^{\mu\nu}$, respectively,  it is straightforward to see that (\ref{eq:g-var0}) can be written as
\begin{equation}\label{eq:g-var1}
\frac{\sqrt{-q}}{\sqrt{-g}}\left(\hat q^{-1}\hat g\right)-\left[\left(\lambda-\frac{\alpha \epsilon}{2} f-\alpha f_R\right)\hat I+\alpha f_R \left(\hat g^{-1}\hat q\right)\right] =-\kappa^2\epsilon \hat {T} \ ,
\end{equation}
where $\hat I$ is the identity matrix, and $\hat{T}$ denotes $T^{\mu\alpha}g_{\alpha\nu}$. This equation establishes an algebraic relation between the object $\hat \Omega\equiv \hat g^{-1}\hat q$ and the matter. In fact, (\ref{eq:g-var1}) can be written as
\begin{equation}\label{eq:g-var2}
|\hat\Omega|^{\frac{1}{2}} \hat \Omega^{-1}-\left[\left(\lambda-\frac{\alpha \epsilon}{2} f-\alpha f_R\right)\hat I+\alpha f_R \hat \Omega\right] =-\kappa^2\epsilon \hat {T} \ .
\end{equation}
Now, multiplying this equation by $\hat \Omega^{-1}$ and defining 
\begin{equation}\label{eq:B}
\hat B=\frac{1}{2|\hat\Omega|^{\frac{1}{2}}}\left[\left(\lambda-\frac{\alpha \epsilon}{2} f-\alpha f_R\right)\hat I-\kappa^2\epsilon \hat {T}\right] \ ,
\end{equation}
we can write (\ref{eq:g-var2}) in the more compact form
\begin{equation}\label{eq:g-var3}
\left(\hat \Omega^{-1}-\hat B\right)^2=\frac{\alpha f_R}{|\hat\Omega|^{\frac{1}{2}}}\hat I +\hat B^2 \ .
\end{equation}
For sources with a diagonal stress-energy tensor, this equation can be solved straightforwardly. Since we are interested in cosmological applications with perfect fluids, we are in one of those simple situations. In the general case, we can formally solve (\ref{eq:g-var3}) in the form 
\begin{equation}
\hat \Omega^{-1}=\hat B\pm \sqrt{\frac{\alpha f_R}{|\hat\Omega|^{\frac{1}{2}}}\hat I +\hat B^2} \ .
\end{equation}
 The sign in front of the square root can be determined by considering the limit to BI theory $\alpha\to 0$. In this case, we get $\lim_{\alpha\to 0} \hat \Omega^{-1}=\lim_{\alpha\to 0}  2\hat B$ if the positive sign is chosen and zero otherwise. Since $\lim_{\alpha\to 0} \hat B=\frac{1}{2|\hat\Omega|^{\frac{1}{2}}}\left[\lambda\hat I-\kappa^2\epsilon \hat {T}\right]$, we find that $\lim_{\alpha\to 0} \hat \Omega^{-1}=\frac{1}{|\hat\Omega|^{\frac{1}{2}}}\left[\lambda\hat I-\kappa^2\epsilon \hat {T}\right]$, which coincides with the corresponding expression found in the literature (see, for instance, \cite{Olmo:2013gqa}).  In the BI case, this last result tells us that $|\hat \Omega|=|\lambda\hat I-\kappa^2\epsilon \hat {T}|$, i.e., $|\hat \Omega|$ is a function of the matter sources and, therefore, $\hat \Omega^{-1}$ is also a function of $\hat T$ and $\lambda$. In our more general scenario, we see that  $|\hat \Omega|$ must depend on $\hat T$ but also on $R$ through the $f(R)$ and $f_R$ terms present in $\hat B$. In principle, for a perfect fluid of  matter density $\rho$ and pressure $P$, with ${T^\mu}_\nu=$diag$[-\rho,P,P,P]$, one can solve for $R$ as a function of the matter using the trace of $\hat \Omega =\hat I +\epsilon \hat R$, where $\hat R$ denotes de matrix $g^{\mu\alpha}R_{\alpha\nu}$, which gives $\Omega^{\mu}_\mu (R, \rho,P)=4 +\epsilon R$. This formally allows to write $R=R(\rho,P)$.  From this, one concludes that $\hat \Omega$ must just be a function of $\rho$ and $P$. In general, though, the explicit dependence of the components of  $\hat \Omega$ on $\rho$ and $P$ might be complicated to obtain and/or may require the use of numerical methods to solve the algebraic relations involved. \\

Having established that $R$ and $\hat \Omega$ can be expressed as functions of the matter, we can now consider the connection equation (\ref{eq:con-var0}), which can also be written as
\begin{equation}\label{eq:con-var1}
\nabla_\beta\left[\sqrt{-g}g^{\mu\lambda}{\Sigma_\lambda}^\nu\right]=0 \ , 
\end{equation}
where we have defined 
\begin{equation}
{\Sigma_\lambda}^\nu\equiv \left(|\hat\Omega|^{\frac{1}{2}}{[\hat \Omega^{-1}]_\lambda}^\nu+\alpha f_R {\delta_\lambda}^\nu\right) \ .
\end{equation}
Given that $\hat \Omega$ is close to the identity except, perhaps, in very extreme high-energy cases, we can assume that ${\Sigma_\lambda}^\nu$ is invertible (at least in some low-energy domain). The invertibility of this object is, however, an assumption that must be verified on a case-by-case basis. Note that, in general, the algebraic dependence of the matrix ${[\hat \Omega^{-1}]_\lambda}^\nu$ on the matter sources cannot be guessed {\it a priori} and depends on the specific $T_{\mu\nu}$ and $f(R)$ considered. For this reason general statements about the invertibility of ${\Sigma_\lambda}^\nu$ at arbitrary energy scales cannot be made without a concrete model.
 Assuming that ${\Sigma_\lambda}^\nu$ is invertible, as will be the case of a perfect fluid to be considered in this work, we can write the term within brackets in the above equation as $\sqrt{-g} \hat{g}^{-1} \hat \Sigma$ and look for an auxiliary metric $\hat{h}$ such that $\sqrt{-g} \hat{g}^{-1} \hat \Sigma=\sqrt{-h} \hat{h}^{-1}$. It is then straightforward to verify that $|g||\hat \Sigma|=|h|$, which implies
\begin{equation}\label{eq:h-g}
\hat h =|\hat\Sigma|^{\frac{1}{2}} \hat \Sigma^{-1} \hat g \ , \ \hat h^{-1} =\frac{1}{|\hat\Sigma|^{\frac{1}{2}}} \hat g^{-1} \hat \Sigma \ .
\end{equation}
The connection equation (\ref{eq:con-var1}) can thus be written as $\nabla_\beta\left[\sqrt{-h}{h}^{\mu\nu}\right]=0$, which implies that $\Gamma^\alpha_{\mu\nu}$ is the Levi-Civita connection of ${h}_{\mu\nu}$. \\

With all these results, we are now ready to write the field equations for the metric in explicit form. Starting from the definition (\ref{eq:q}), and knowing that $\Gamma^\alpha_{\mu\nu}$ is the Levi-Civita connection of $h_{\mu\nu}$, we have that $R_{\mu\nu}(\Gamma)=R_{\mu\nu}(h)=(q_{\mu\nu}-g_{\mu\nu})/\epsilon$. Raising one index of this equation with $h^{\nu\alpha}$ and using the definitions of $\hat \Sigma$ and $\hat \Omega$, we get
\begin{equation}\label{eq:Rmn-h}
{R_\mu}^{\beta} (h)=\frac{{\Sigma_\mu}^{\gamma}}{\epsilon {|\hat{\Sigma}|}^{\frac{1}{2}}}\left[{\Omega_\gamma}^\beta- {\delta_\gamma}^{\beta}\right]
\end{equation}
We remark that both ${\Sigma_\mu}^{\gamma}$ and ${\Omega_\mu}^{\gamma}$ are functions of the matter. Therefore, the sources appear on the right-hand side of this equation, whereas the left-hand side contains derivatives of $h_{\mu\nu}$ up to second-order. One can thus solve the equations for $h_{\mu\nu}$ and then use the relations (\ref{eq:h-g}) to obtain $g_{\mu\nu}$. \\

We now discuss the field equations in vacuum. When $\hat T$ vanishes, we find that  $\hat B$, $\hat\Omega$, and $\hat \Sigma$ are proportional to the identity. The trace of $\hat \Omega$ can be used to show that $R$ must be a constant, whose value depends on the particular form of the model chosen. As a result, we find that $h_{\mu\nu}=C g_{\mu\nu}$, where $C$ is a constant factor, and (\ref{eq:Rmn-h}) boils down to ${R_\mu}^\nu (h)=\tilde{C} {\delta_\mu}^\nu$, which is equivalent to the vacuum field equations of GR+$\Lambda$, namely, $R_{\mu\nu}(g)=\Lambda g_{\mu\nu}$. This result puts forward that a very large family of gravity theories formulated within the Palatini approach yield the same vacuum dynamics as GR, though they differ in those regions where the energy-density is nonzero. Einstein's equations, therefore, appear as a very fundamental property of metric-affine (Palatini) theories of gravity \cite{Mauro1,Mauro2}. \\

\section{Perfect fluid scenarios \label{sec:fluids}}

For a perfect fluid with energy-density $\rho$, pressure $P$, and stress-energy tensor of the form $T_{\mu\nu}=(\rho+P)u_\mu u_\nu+P g_{\mu\nu}$, we find that 
\begin{equation}\label{eq:Bpf}
{B_\mu}^\nu=\frac{1}{2|\hat\Omega|^{\frac{1}{2}}}\left(\begin{array}{cc}
b_1 & \vec{0} \\
\vec{0} & b_2 \hat I_{3\times 3}
\end{array}\right) \ ,
\end{equation}
where 
\begin{eqnarray}\label{eq:b0}
b_1&\equiv& \left[\lambda-\alpha\left(\epsilon f/2+f_R\right)+\epsilon\kappa^2 \rho\right] \\
b_2&\equiv & \left[\lambda-\alpha\left(\epsilon f/2+f_R\right)-\epsilon\kappa^2 P\right]\label{eq:b1}
\end{eqnarray}
 With this one immediately finds that
\begin{eqnarray}
{\Omega_\mu}^\nu&=&{2|\hat\Omega|^{\frac{1}{2}}}\left(\begin{array}{cc}
w_1 & \vec{0} \\
\vec{0} & w_2\hat I_{3\times 3}
\end{array}\right) \\
 {[\hat \Omega^{-1}]_\mu}^\nu&=&\frac{1}{2|\hat\Omega|^{\frac{1}{2}}}\left(\begin{array}{cc} w_1^{-1}
 & \vec{0} \\
\vec{0} &w_2^{-1} \hat I_{3\times 3}
\end{array}\right)  \\
w_i&\equiv& 
\left[b_i+\sqrt{b_i^2+4\alpha f_R|\hat\Omega|^{\frac{1}{2}}}\right]^{-1} \label{eq:wi} \ .
\end{eqnarray}
Note that in the definition of $w_i$ there appears a square root and, therefore, some positivity conditions must be satisfied for the consistency of the model. 
In this sense, when the $f(R)$ term is negligible, or in the limit $\alpha\to 0$, we get $w_i=2b_i$, which recovers the result of the BI case. If the $\alpha f_R$ term becomes negative, as will be our case, the square root may become zero or even reach negative values, which could lead to inconsistencies. As we will see in more detail later, the dynamics prevents such pathological situations. \\

The determinant of $\hat \Omega^{-1}$ leads to 
\begin{equation}\label{eq:det}
16{|\hat\Omega|}=1/(w_1w_2^{3}) \ ,
\end{equation}
whereas the trace of $\hat \Omega$ yields
\begin{equation}\label{eq:trace}
4+\epsilon R={2|\hat\Omega|^{\frac{1}{2}}}\left(w_1+3w_2\right) \ .
\end{equation}
Combining (\ref{eq:det}) and (\ref{eq:trace}) one should be able, in principle, to obtain expressions for $R$ and $|\hat\Omega|$ in terms of $\rho$ and $P$.  \\

\subsection{General expressions for $\rho$ and $P$}

We mentioned above that (\ref{eq:det}) and (\ref{eq:trace}) establish algebraic relations between the variables $\rho$, $P$, $R$,  and $|\hat\Omega|$, in such a way that only two of them are actually independent. The most satisfactory case is that in which $R$ and $|\hat\Omega|$ can be explicitly written in terms of $\rho$ and $P$. In general, however, the situation could be nontrivial and numerical methods might be necessary to establish that relation, but this is just a technical question. In this sense, it is relatively straightforward to find an expression for $\rho$ and $P$ in terms of  $R$,  and $|\hat\Omega|$ without the need for specifying the particular $f(R)$ Lagrangian. This approach yields $\rho$ and $P$ in parametric form. 

The idea is to start from  (\ref{eq:det}) and write it in the form 
\begin{equation}\label{eq:det1}
\frac{1}{\left[b_1+\sqrt{b_1^2+4\alpha f_R|\hat\Omega|^{\frac{1}{2}}}\right]}=\frac{\left[b_2+\sqrt{b_2^2+4\alpha f_R|\hat\Omega|^{\frac{1}{2}}}\right]^3}{16{|\hat\Omega|}} \ .
\end{equation}
This relation can be inserted in (\ref{eq:trace}) to remove the dependence on $\rho$ or to remove the dependence on $P$ (recall from the definitions (\ref{eq:b0}) and (\ref{eq:b1}) that $b_1$ depends on $R$ and $\rho$ whereas $b_2$ depends on $R$ and $P$). For instance, using (\ref{eq:det1}) to remove the dependence on $\rho$ from (\ref{eq:trace}) and defining $\delta_2\equiv \left[b_2+\sqrt{b_2^2+4\alpha f_R|\hat\Omega|^{\frac{1}{2}}}\right]$, we get 
\begin{equation}\label{eq:trace1}
4+\epsilon R={2|\hat\Omega|^{\frac{1}{2}}}\left(\frac{\delta_2^3}{16{|\hat\Omega|}}+\frac{3}{\delta_2}\right) \ .
\end{equation}
From this one can obtain an expression for $\delta_2$ in terms of $R$ and $|\hat\Omega|$ by just finding the roots of a quartic polynomial, which can be carried out with the use of tables or algebraic manipulation software. The following step consists on inverting the relation between $b_2$ and $\delta_2=\delta_2(R,|\hat\Omega|)$, which allows to write $P$ as 
\begin{equation}\label{eq:P}
\epsilon\kappa^2 P= \lambda-\alpha\left(\epsilon f+f_R\right)-\frac{\delta_2^2-4\alpha f_R |\hat\Omega|^{\frac{1}{2}}}{2\delta_2} \ .
\end{equation}
A similar approach can be used to extract $\rho$ from $\delta_1\equiv \left[b_1+\sqrt{b_1^2+4\alpha f_R|\hat\Omega|^{\frac{1}{2}}}\right]$. In this case, one gets 
\begin{equation}\label{eq:trace2}
4+\epsilon R={2|\hat\Omega|^{\frac{1}{2}}}\left(\frac{1}{\delta_1}+\frac{3\delta_1^{\frac{1}{3}}}{(4x)^\frac{2}{3}}\right) \ ,
\end{equation}
which becomes a quartic equation for the variable $\gamma\equiv \delta_1^{-\frac{1}{3}}$.  The procedure is analogous to the previous case and yields
\begin{equation}\label{eq:rho}
\epsilon\kappa^2 \rho=-\left[ \lambda-\alpha\left(\epsilon f+f_R\right)\right]+\frac{\delta_1^2-4\alpha f_R |\hat\Omega|^{\frac{1}{2}}}{2\delta_1} \ .
\end{equation}

\section{Cosmology \label{sec:cosmology}}

In order to study the cosmology of the BI$-f(R)$ family of models introduced in the previous sections, we must find first an expression for the Hubble function. To proceed, we consider an homogeneous and isotropic Friedman-Lemaitre-Robertson-Walker (FLRW) line element in the spatially flat case,
\begin{equation}
ds^2=g_{\mu\nu}dx^\mu dx^\nu=-dt^2+a^2(t)\delta_{ij}dx^i dx^j \ ,
\end{equation}
and use relations (\ref{eq:h-g}) to find its relation with the components of $h_{\mu\nu}$ necessary to use the field equations (\ref{eq:Rmn-h}). Following a notation similar to that used in \cite{Olmo_book}, we can write 
\begin{equation}
{\Sigma_\mu}^\nu=\left(\begin{array}{cc}
\sigma_1 &  \vec{0} \\
\vec{0} & \sigma_2 \hat{I}_{3\times 3}
\end{array}\right) \ , \ \sigma_i=\alpha f_R+\frac{\delta_i}{2} \ , \ i=1,2 \ ,
\end{equation}
which implies 
\begin{eqnarray}
h_{tt}&=& -\sqrt{\frac{ \sigma_2^3}{\sigma_1}}\\
h_{ij}&=& \sqrt{\sigma_1 \sigma_2}a^2(t)\delta_{ij} \equiv \Delta(t) a^2(t)\delta_{ij}\ .
\end{eqnarray}
Recall that since $\sigma_1$ and $\sigma_2$ are functions of $\rho$ and $P$, it follows that  $\Delta$ is a function of time, as we have explicitly written above. This is the only aspect we need to know so far to proceed with the derivation of the Hubble equation. After a bit of algebra, one gets
\begin{equation}\label{eq:Gtt}
G_{tt}\equiv 3\left(H+\frac{\dot\Delta}{2\Delta}\right)^2 \ ,
\end{equation}
where $H\equiv \dot{a}/a$. From the field equation (\ref{eq:Rmn-h}), we find that 
\begin{equation}\label{eq:Gtt-rhs}
\epsilon G_{tt}= \frac{\sigma_1-3\sigma_2 -2|\hat\Omega|^{\frac{1}{2}}(\sigma_1 w_1-3\sigma_2 w_2)}{2\sigma_1} \ ,
\end{equation}
which in combination with (\ref{eq:Gtt}) yields
\begin{equation}\label{eq:Hubble}
3\epsilon\left(H+\frac{\dot\Delta}{2\Delta}\right)^2= \frac{\sigma_1-3\sigma_2 -2|\hat\Omega|^{\frac{1}{2}}(\sigma_1 w_1-3\sigma_2 w_2)}{2\sigma_1} \ .
\end{equation}
For a  fluid with equation of state $\omega=P/\rho$, we have that $\Delta=\Delta(\rho, \omega)$ and, therefore, $\dot\Delta=\Delta_{\rho}\dot\rho$, where $\Delta_{\rho}\equiv \partial\Delta/\partial {\rho}$. Since the conservation equation is  $\dot\rho=-3H (1+\omega)\rho$, we find that $\dot\Delta=-3H  (1+\omega) \rho \Delta_{\rho}$. With this result, (\ref{eq:Hubble}) leads to 
\begin{equation}\label{eq:Hubble-pfluids}
\epsilon H^2=\frac{\sigma_1-3\sigma_2 -2|\hat\Omega|^{\frac{1}{2}}(\sigma_1 w_1-3\sigma_2 w_2)}{2\sigma_1\left(1-\frac{3  (1+\omega) \rho \Delta_{\rho}}{2\Delta}\right)^2} \ ,
\end{equation}
Note that all the quantities appearing on the right-hand side of this equation are functions of the matter density $\rho$, which allows to obtain a parametric representation of $H^2$ as a function of $\rho$. This can be used, in particular, to determine if for a given choice of $f(R)$ and equation of state $\omega$ bouncing solutions exist.

\subsection{A model $f(R)=R^2$}

To illustrate the procedure to deal with the theories presented in this work, we consider a simple model characterized by a function $f(R)=R^2$. This model can be treated analytically and allows to modify the coefficient in front of the $R^2$ term that arises in the original Born-Infeld gravity theory. In fact, a series expansion of the Born-Infeld action for small values of the parameter $\epsilon$ leads to \begin{equation}
\lim_{\epsilon \rightarrow 0} S = \frac{1}{2\kappa^2} \int d^4x \sqrt{-g} \left[R-2\Lambda_{eff}+\frac{\epsilon R^2}{4} -\frac{\epsilon}{2} R_{\mu\nu}R^{\mu\nu}+\ldots+\alpha f(R) \right] +S_m \label{eq:BI_series} \ ,
\end{equation}
where  $\Lambda_{eff}=\frac{\lambda -1}{\epsilon}$. The coefficients in front of the quadratic (and all higher-order) curvature terms coming from the BI action are fixed. However, by adding an $f(R)$ piece to the Lagrangian, we can vary the $R-$dependent terms at will. For illustration purposes, we consider $\alpha f(R)=-a\epsilon R^2/4$, which for $a=0$ recovers the original BI theory whereas for $a=1$ completely cancels out the $R^2$ contribution.  \\

 \begin{figure}[h]
\begin{center}
\includegraphics[width=0.75\textwidth]{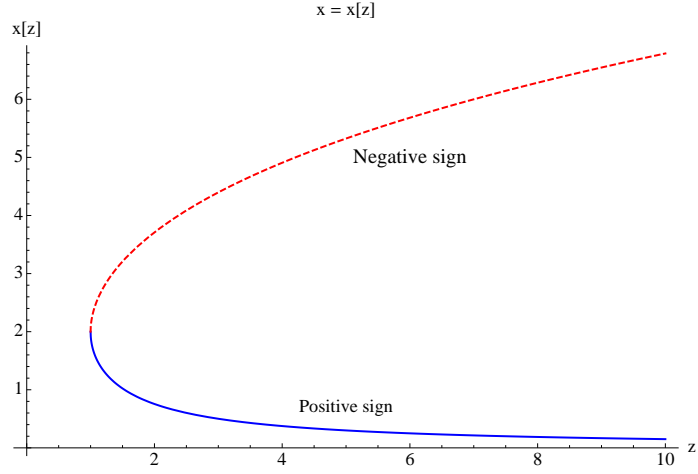}
\caption{Representation of the two branches of the function $x(z)$ defined in Eq.(\ref{eq:x(z)}) (which is identical with $y(z)$) depending on the sign in front of the square root. \label{fig:1}} 
\end{center}
\end{figure}

In order to determine the impact of changing the coefficient in front of the $R^2$ piece from the above action on the Hubble function (\ref{eq:Hubble-pfluids}), we need to work out the dependence of $P$ and $\rho$ on $R$ and $|\hat\Omega|$ using formulas (\ref{eq:P}) and (\ref{eq:rho}). The first step is to solve $\delta_2$ from (\ref{eq:trace1}). To do it, it is convenient to introduce the redefinition $\delta_2=x|\hat\Omega|^{1/4}$, which turns  (\ref{eq:trace1}) into 
\begin{eqnarray}
2z=\frac{x^3}{16}+\frac{3}{x}  \\
z\equiv \frac{4+\epsilon R}{4|\hat\Omega|^{1/4}} \ .
\end{eqnarray}
This equation, which is independent of the $f(R)$ theory considered, admits the physical solutions (see Fig. \ref{fig:1} for a graphic representation of $x$)
\begin{eqnarray}\label{eq:x(z)}
x&=&\frac{\sqrt{2} \left(\Phi ^{3/4} \pm \sqrt{2^{3/2} z-\Phi ^{3/2}}\right)}{{\Phi }^{1/4}} \\
\Phi &=&\left({z^2-\sqrt{z^4-1}}\right)^{1/3}+\left({\sqrt{z^4-1}+z^2}\right)^{1/3}
\end{eqnarray}
With this result, one finds that (\ref{eq:P}) can be written as 
\begin{equation}\label{eq:P2}
\epsilon\kappa^2 P= \lambda-\alpha\left(\epsilon f/2+f_R\right)-\frac{|\hat\Omega|^{\frac{1}{4}}}{2}\frac{(x^2-4\alpha f_R) }{x} \ .
\end{equation}
The equation for $\rho$ can be manipulated in a very similar way. Introducing the replacement $\delta_1=16 |\hat\Omega|^{\frac{1}{4}}/y^3$, (\ref{eq:trace2}) becomes
\begin{equation}
2z=\frac{y^3}{16}+\frac{3}{y}  \ ,  
\end{equation} 
which admits the same solution as $x$. As we will see later, the existence of two possible signs in the definitions of $x$ and $y$ must be taken into account for the correct identification of the physical solutions. With this result, one finds that (\ref{eq:rho}) can be written as 
\begin{equation}\label{eq:rho2}
\epsilon\kappa^2 \rho= -[\lambda-\alpha\left(\epsilon f/2+f_R\right)]+\frac{|\hat\Omega|^{\frac{1}{4}}}{8}\frac{(64-\alpha f_Ry^6) }{y^3} \ ,
\end{equation}
where 
\begin{eqnarray}
y&=&\frac{\sqrt{2} \left(\Phi ^{3/4} \pm \sqrt{2^{3/2} z-\Phi ^{3/2}}\right)}{{\Phi }^{1/4}} \\
\Phi &=&\left({z^2-\sqrt{z^4-1}}\right)^{1/3}+\left({\sqrt{z^4-1}+z^2}\right)^{1/3}
\end{eqnarray}

One can verify that with the definitions of $x$ and $y$ given here, we must have $z\ge1$.  On the other hand, once a value of $z$ is set, the definition of $z$ implies a relation between $\epsilon R$ and $|\hat\Omega|^{\frac{1}{4}}$, which means that only two variables are needed to parametrize the functions $P$ and $\rho$. In the case of a perfect fluid with equation of state $\omega=P/\rho=$constant, a relation between the two independent variables arises and only one variable is needed. In fact, for constant $\omega$ we find 
\begin{equation}\label{eq:R_z}
|\hat\Omega|^{\frac{1}{4}}=\frac{2(1+\omega)[\lambda-\alpha(\epsilon f/2+f_R)]}{\frac{x^2-4\alpha f_R}{x}+\frac{\omega}{4}\frac{64-\alpha f_R y^6}{y^3}} \ .
\end{equation}
Now, since $|\hat\Omega|^{\frac{1}{4}}=(4+\epsilon R)/(4z)$, (\ref{eq:R_z}) establishes a relation between $R$ and $z$, which at the same time allows us to write $|\hat\Omega|$ as a function of $z$.  \\

To illustrate this point, consider  the case $\alpha f(R)=-a\epsilon R^2/4$, which interpolates between the BI theory ($a=0$) and the BI$-f(R)$ case without $R^2$ term ($a=1$). Though an exact expression for arbitrary $\omega$ can be found, for  $\omega=0$ it simplifies to 
\begin{equation}\label{eq:R}
\epsilon R(z)=\frac{x^2+a (8-4 x z)\pm  \sqrt{16 a^2 (x z-2)^2+8 a x \left(x^2 z-4 x z^2-2 x+8 z\right)+x^4}}{2 a (x z-2)} \ ,
\end{equation}
which is valid for any $a\neq 0$. In order to have a well-defined limit to BI theory as $a\to 0$, one must take the minus sign in front of the square root. In that case, the divergent term of the above expression as $a\to 0$ vanishes and we find that $\epsilon R(z)_{BI}$ is given by the zeroth-order term in a series expansion in the parameter $a$ (the formula given here is also valid for arbitrary $\omega$):
\begin{equation}\label{eq:R}
\epsilon R(z)_{BI}=\frac{4 y^3 (2 (\omega +1) z-x)-64 \omega }{\left(x y^3+16 \omega \right)} \ .
\end{equation}
It is important to note that both $x=x(z)$ and $y=y(z)$ have two possible signs each. The right choice must be determined on physical grounds, as we will see shortly.

\begin{figure}[h]
\includegraphics[width=0.75\textwidth]{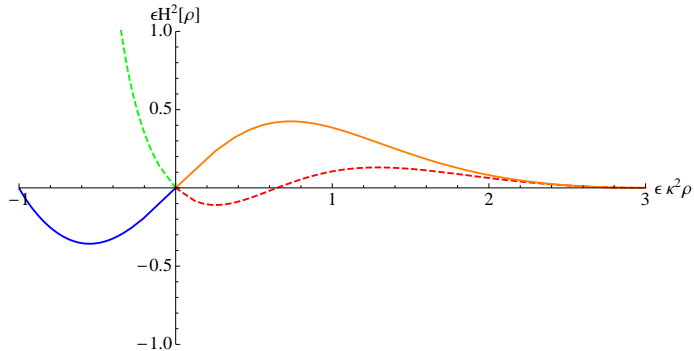}
\caption{Representation of the (dimensionless) Hubble function $\epsilon H^2$ as a function of the (dimensionless) energy density $\epsilon \kappa^2\rho$ in the original BI theory with equation of state $w=1/3$ for the different combinations of signs in the functions $x(z)$ and $y(z)$. The solid orange curve contained in the upper right quadrant represents the $(-,-)$ solution. The dashed green curve contained in the upper left quadrant represents the $(-,+)$ solution. The solid blue curve contained in the lower left quadrant represents the $(+,+)$ solution. The other dashed curve is the $(+,-)$ case. Note that the $(+,+)$ solution becomes physical ($\epsilon H^2>0$ and  $\epsilon \kappa^2\rho>0$) if $\epsilon <0$.   \label{Fig:H2BI_rad}} 
\end{figure}

\subsubsection{Hubble function}

With the above expressions for $\epsilon R(z)$ (and their generalization to arbitrary $\omega$), one can completely parametrize $|\hat\Omega|$, $\epsilon\rho$,  $\epsilon P$, and $\epsilon H^2$ in terms of $z$. This allows us to obtain graphic representations of $\epsilon H^2$ as a function of $\epsilon\rho$, which can be used to study the nature and robustness of the zeros of the Hubble function at high densities as the parameters of the theory are modified.  \\

Let us consider first the original BI theory. The parametrization in terms of the variable $z$ given above yields four solutions that represent the possible combinations of signs in the functions $x$ and $y$. From the plot shown in Fig. \ref{Fig:H2BI_rad}, which represents the case $w=1/3$ (a universe filled with radiation), it is clear that only the $(+,+)$ and $(-,-)$ solutions are physical, since the other two represent either a case with positive $\epsilon\rho$ but negative $\epsilon H^2$ or positive $\epsilon H^2$ with negative energy density $\epsilon\rho$. A similar behavior is also observed in the BI$-f(R)$ case (not shown in the plot).

\begin{figure}[h]
\begin{center}
\includegraphics[width=0.75\textwidth]{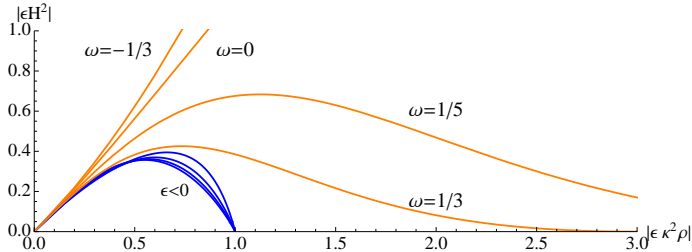}
\caption{Representation of the (dimensionless) Hubble function $\epsilon H^2$ as a function of the (dimensionless) energy density $\epsilon \kappa^2\rho$ in the original BI theory for different equations of state ($w=-1/3,0,1/5,$ and $1/3$).  \label{Fig:H2BI_w}}
\end{center}
\end{figure}

In Fig.  \ref{Fig:H2BI_w} we see that for those solutions with $\epsilon <0$ the Hubble function vanishes at $|\epsilon \kappa^2\rho|=1$ regardless of the sign of $w$. These solutions represent a cosmic bounce characterized by $H^2=0$ and $ dH^2/d\rho\neq 0$. The behavior of $|\epsilon H^2|$ for $\epsilon>0$ is more sensitive to the value of $w$, having a divergent behavior for $w\le 0$. For $\omega>0$, $H^2$ vanishes at a finite density $\epsilon\kappa^2\rho_c=1/\omega$. These solutions do not represent a cosmic bounce, but an unstable state of minimum volume  \cite{Banados} characterized by $H^2=0=dH^2/d\rho$. \\

\begin{figure}[h]
\begin{center}
\includegraphics[width=0.65\textwidth]{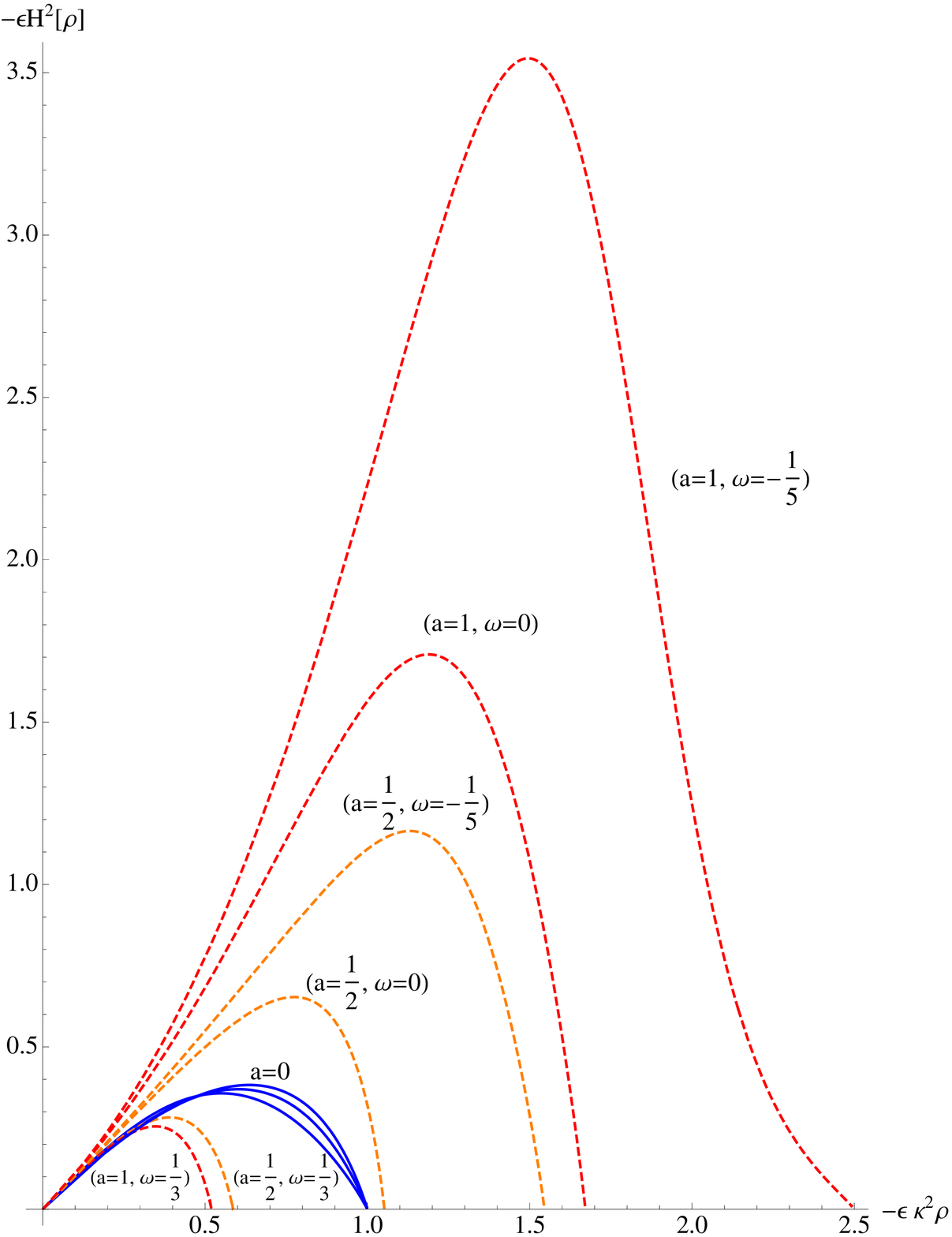}
\caption{Representation of the (dimensionless) Hubble function $-\epsilon H^2$ as a function of the (dimensionless) energy density $-\epsilon \kappa^2\rho$ in the original BI theory (solid blue) and in two quadratic modifications of the form $f(R)=a R^2$, with $a=1/2$ (dashed orange) and $a=1$ (dashed red), for different equations of state ($w=-1/5,0,$ and $1/3$). The existence of a bounce appears as a robust property of the $\epsilon<0$ branch of the theory. \label{Fig:Negative_Branch}}
\end{center}
\end{figure}

When the coefficient of the $R^2$ term is modified, the existence of a cosmic bounce appears as a robust property of the $\epsilon<0$ branch of the theory (see Fig. \ref{Fig:Negative_Branch}). The $\epsilon>0$ branch, on the contrary, exhibits a strong sensitivity to variations in the $R^2$ term. In fact, in the lower right plot of  Fig. \ref{Fig:Positive_Branch}, we see that the loitering behavior of the radiation universe observed in the BI theory is highly unstable and disappears as we move away from the original BI case.  It should be noted, however, that other similar stationary points arise for equations of state $\omega \lesssim 1/10$ and persist even for negative values of $\omega$, which contrasts with the BI theory. \\

It is worth noting that, as shown in the lower left plot of Fig. \ref{Fig:Positive_Branch},   after a local maximum $H^2$ may reach a non-zero minimum followed by a divergence at a large finite value of the energy density. Though these solutions do not avoid the big bang singularity, they possess another very interesting property, namely, the existence of a long plateau comprised between a local minimum and a local maximum that appears at lower energies. This plateau on $H^2$ may naturally yield a period of approximately de Sitter cosmic inflation shortly after the big bang. In Fig.\ref{Fig:Inflation} we illustrate this property also in a radiation universe with $a=1/3$ (green dashed curve).

\begin{figure}[h]
\begin{center}
\includegraphics[width=1\textwidth]{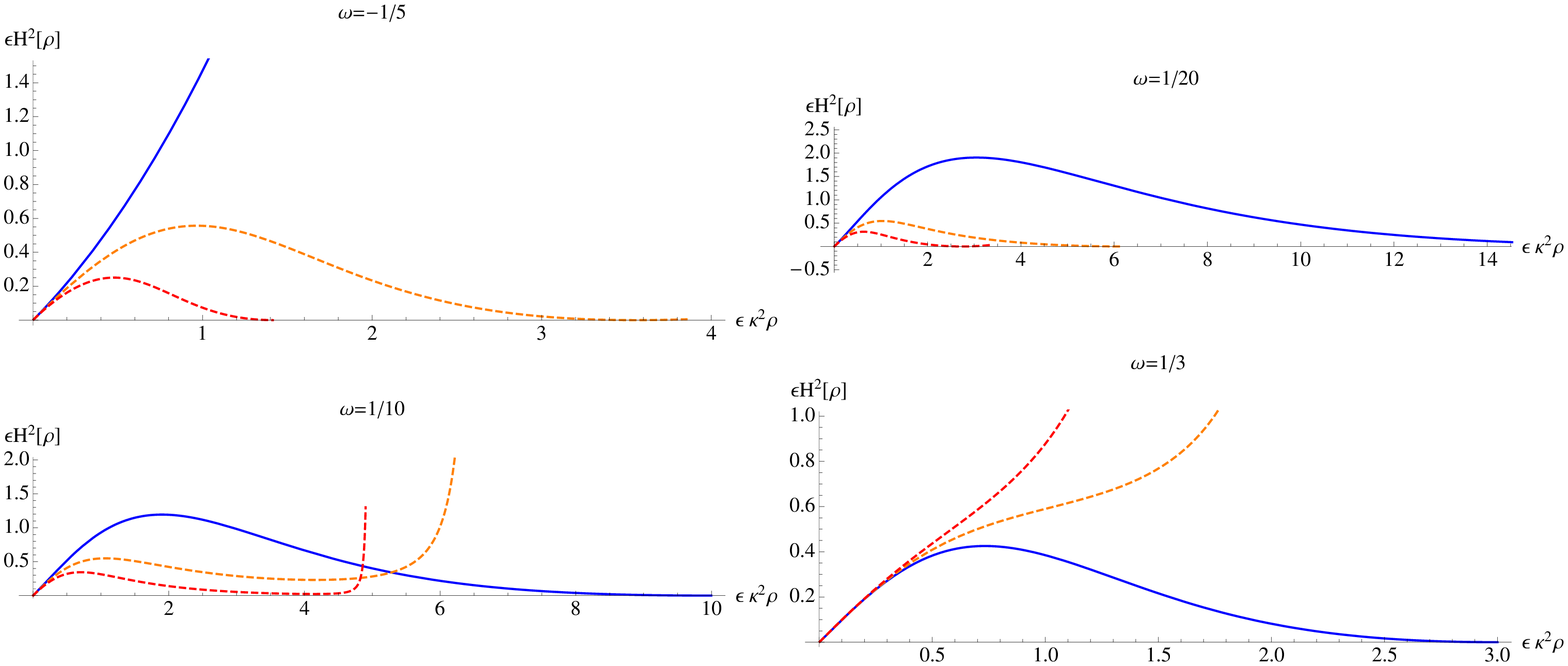}
\caption{Representation of the (dimensionless) Hubble function $\epsilon H^2$ as a function of  the (dimensionless) energy density $\epsilon \kappa^2\rho$ in the original BI theory (solid blue) and in two quadratic modifications of the form $f(R)=a R^2$, with $a=1/2$ (dashed orange) and $a=1$ (dashed red), for different equations of state ($w=-1/5,1/20, 1/10,$ and $1/3$). The zero of  $\epsilon H^2$ for the radiation universe ($\omega=1/3$) is unstable under changes of the parameter $a$ (recall that BI corresponds to $a=0$). As the equation of state approaches $\omega\to 0$, we find that $\epsilon H^2$ may become again zero at high densities. At this point, one can verify that the function $\epsilon\dot{H}$ has a zero, thus implying a minimum of $\epsilon H^2$. This signals an instability representing a state of minimum volume that is not a bounce.  
 \label{Fig:Positive_Branch}}
\end{center}
\end{figure}

Before concluding, let us comment on a technical aspect related with the nature of the solutions presented here. Since we are considering an $f(R)$ model with $4\alpha f_R =-2a\epsilon R$, one might wonder what happens to the square root of the $w_i$ terms in (\ref{eq:wi}) and to $|\hat\Omega|$ at high energies. To illustrate this point, in Fig.\ref{Fig:H2a1} we have plotted the Hubble function corresponding to the case $a=1$ for different equations of state. We have included here both the bouncing solutions of the $\epsilon<0$ branch (dashed curves) and also the solutions of the $\epsilon>0$ branch which, in general, possess a big bang singularity (continuous curves). At low energies, where the GR regime dominates, we find $|\hat\Omega|\sim 1$ for all equations of state (see Fig.\ref{Fig:Det}). At higher energies, the behavior for the BI-$f(R)$ theory is clearly dependent on the particular equation of state and the sign of $\epsilon$. For the original BI theory, however, the behavior is quite generic and only depends on the sign of $\epsilon$ (see the green dotted lines). The green lines form a large finite angle when they cut the density axis at $\epsilon\rho_B=-1$ (bouncing solutions), whereas the angle tends to zero on the right-hand side (unstable, finite volume solutions). In the BI-$f(R)$ case, we see that $|\hat\Omega|$ for the bouncing solutions (dashed lines on the left quadrant) never vanishes and, in fact, is not defined beyond a certain point, which determines the maximum density attained at the bounce. For the $\epsilon>0$ solutions, we see that   $|\hat\Omega|$ can vanish at a certain, finite high energy density, which defines the density at which the Hubble function diverges. Note also that for $\omega=0$ we can have $H^2\to 0$ with nonzero $|\hat\Omega|$, which indicates that nonsingular (possibly unstable) solutions exist for $\epsilon>0$. \\
The terms within the square root of the functions $w_i$ have been denoted as $R_i\equiv b_i^2+4\alpha f_R |\hat\Omega|^{1/2}$ and plotted in Figs.\ref{Fig:R1} and \ref{Fig:R2}. These functions are well behaved over all their physical domain of definition. For completeness, the curvature $R$ for these cases has also been represented in Fig.\ref{Fig:Curv} and compared with the prediction from the original BI theory. 

\begin{figure}[h]
\begin{center}
\includegraphics[width=1\textwidth]{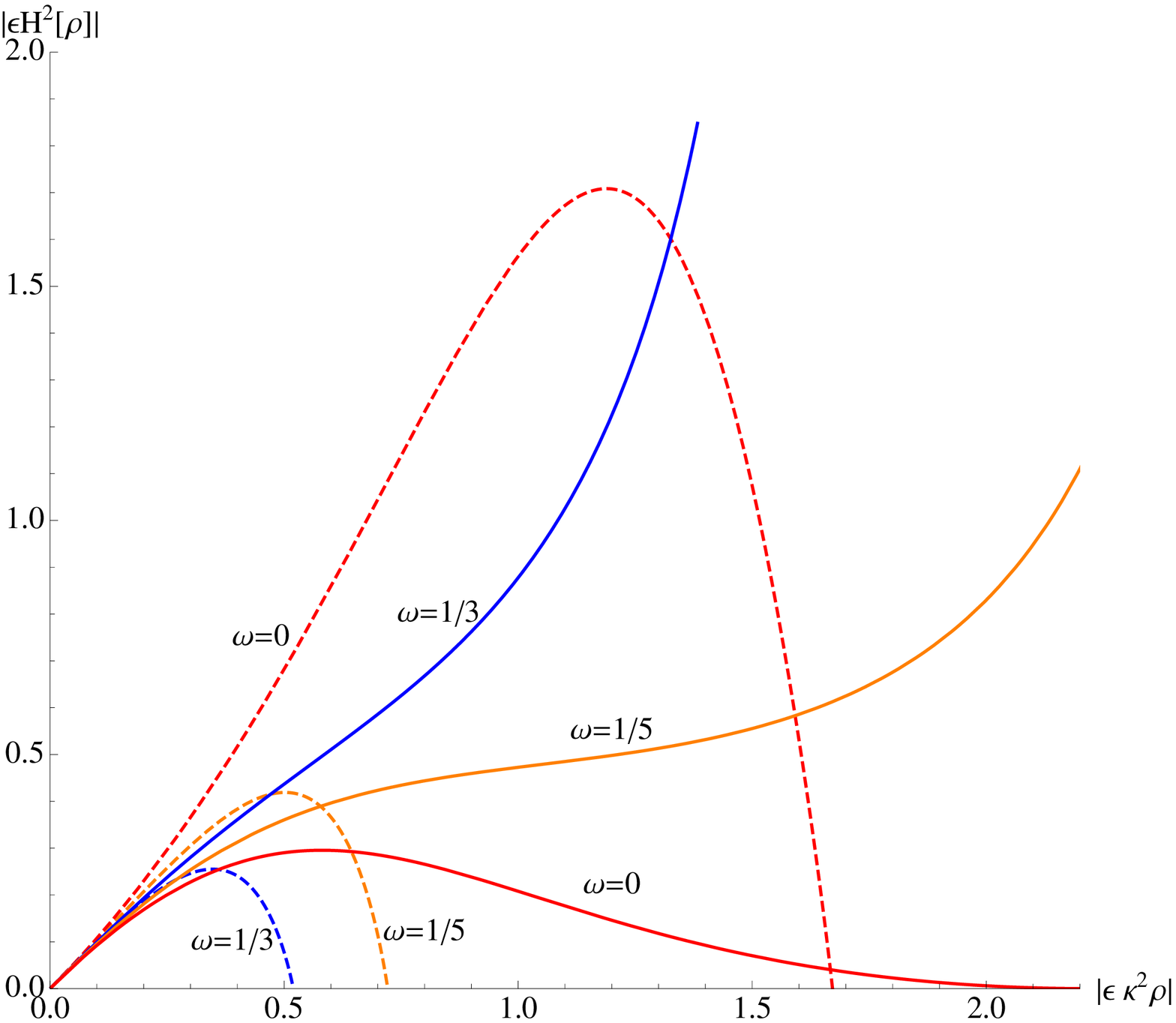}
\caption{Representation of the (dimensionless) Hubble function $\epsilon H^2$ as a function of the (dimensionless) energy density $\epsilon \kappa^2\rho$ for the BI-$f(R)$ theory with $\alpha f(R)= -a\epsilon R^2/4$ and $a=1$. The dashed curves represent bouncing solutions of the branch $\epsilon<0$. The solid curves correspond to the branch $\epsilon>0$.   The equations of state represent dust (red), radiation (blue), and a fluid with $\omega=1/5$ (orange). 
 \label{Fig:H2a1}}
\end{center}
\end{figure}

\begin{figure}[h]
\begin{center}
\includegraphics[width=1\textwidth]{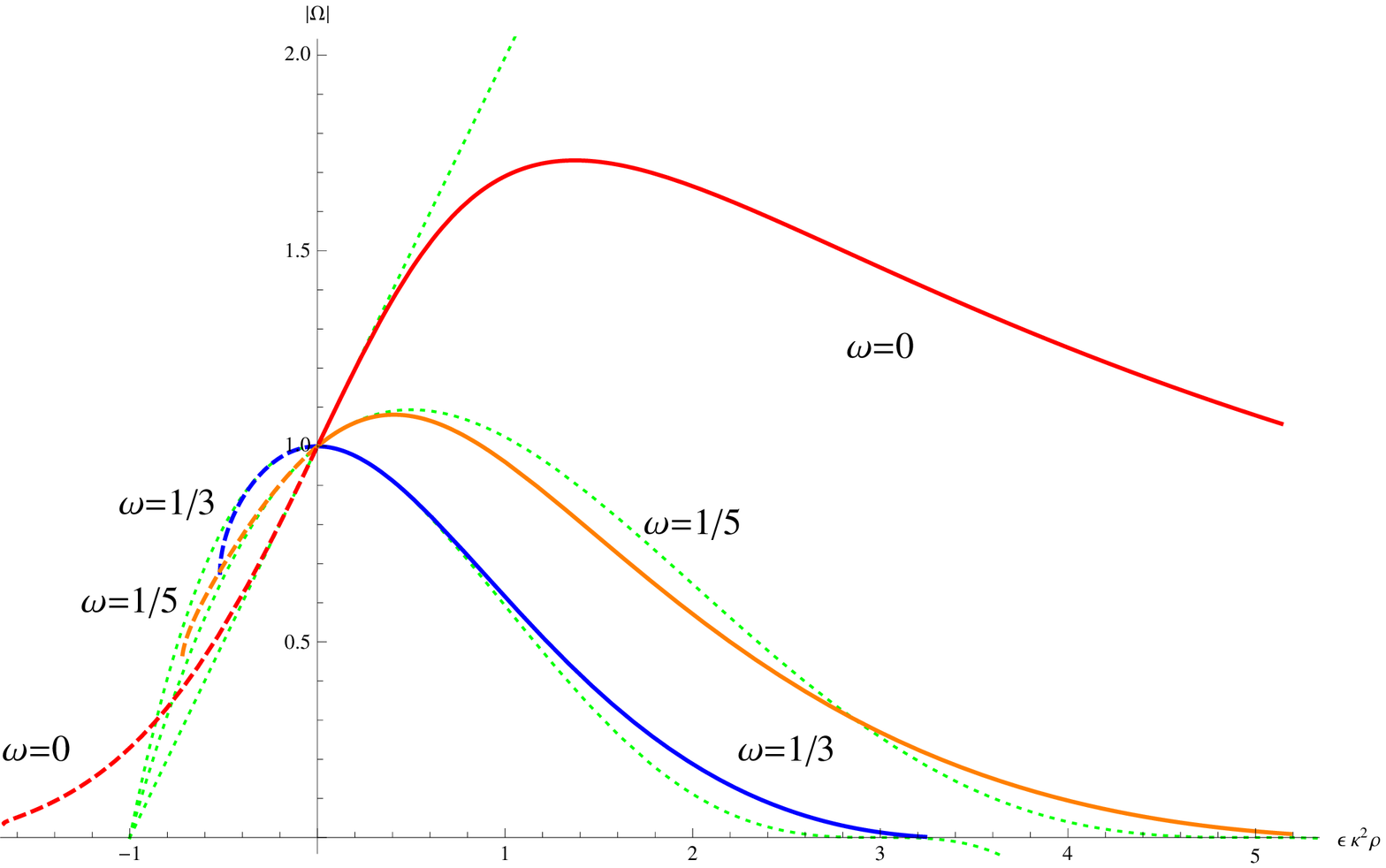}
\caption{Representation of the determinant $|\hat\Omega|$ as a function of the (dimensionless) energy density $\epsilon \kappa^2\rho$ for the BI-$f(R)$ theory with $\alpha f(R)= -a\epsilon R^2/4$ and $a=1$. The dashed curves represent bouncing solutions of the branch $\epsilon<0$. The solid curves correspond to the branch $\epsilon>0$.   The equations of state represent dust (red), radiation (blue), and a fluid with $\omega=1/5$ (orange). The green dotted lines represent the corresponding solutions in the original BI theory. 
 \label{Fig:Det}}
\end{center}
\end{figure}

\begin{figure}[h]
\begin{center}
\includegraphics[width=1\textwidth]{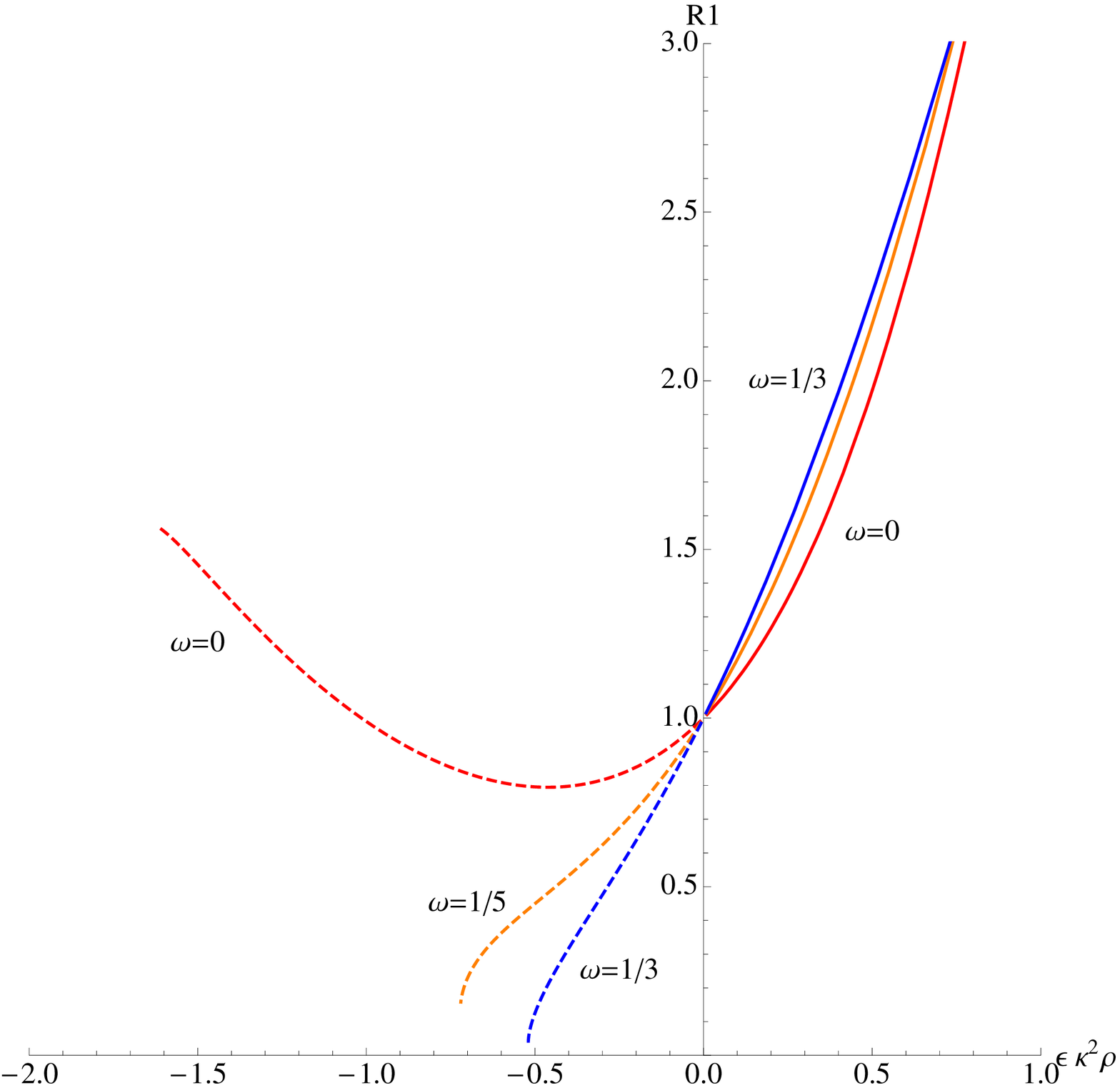}
\caption{Representation of the function $R_1=b_1^2+4\alpha f_R |\hat\Omega|^{1/2}$, which appears under the square root in (\ref{eq:wi}), as a function of the (dimensionless) energy density $\epsilon \kappa^2\rho$ for the BI-$f(R)$ theory with $\alpha f(R)= -a\epsilon R^2/4$ and $a=1$. The dashed curves correspond to the bouncing solutions of the branch $\epsilon<0$. The solid curves correspond to the branch $\epsilon>0$. The equations of state represent dust (red), radiation (blue), and a fluid with $\omega=1/5$ (orange).
 \label{Fig:R1}}
\end{center}
\end{figure}

\begin{figure}[h]
\begin{center}
\includegraphics[width=1\textwidth]{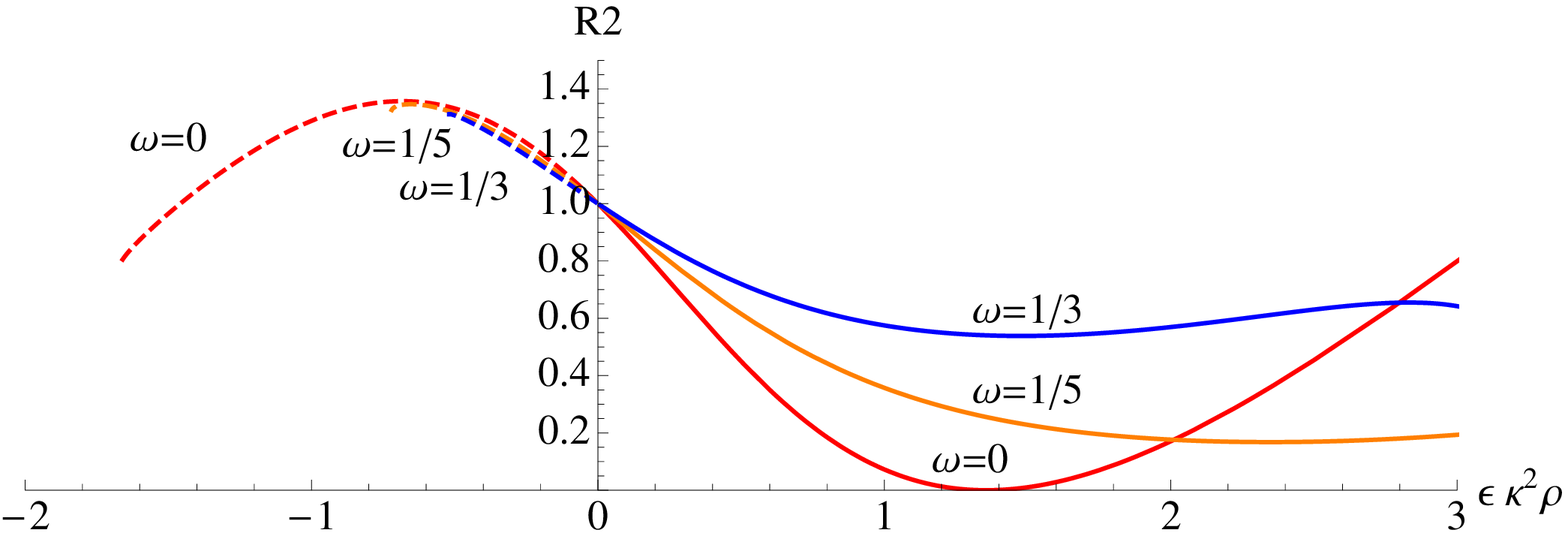}
\caption{Representation of the function $R_2=b_2^2+4\alpha f_R |\hat\Omega|^{1/2}$, which appears under the square root in (\ref{eq:wi}), as a function of the (dimensionless) energy density  $\epsilon \kappa^2\rho$ for the BI-$f(R)$ theory with $\alpha f(R)= -a\epsilon R^2/4$ and $a=1$. The dashed curves correspond to the bouncing solutions of the branch $\epsilon<0$. The solid curves correspond to the branch $\epsilon>0$. The equations of state represent dust (red), radiation (blue), and a fluid with $\omega=1/5$ (orange).
 \label{Fig:R2}}
\end{center}
\end{figure}

\begin{figure}[h]
\begin{center}
\includegraphics[width=1\textwidth]{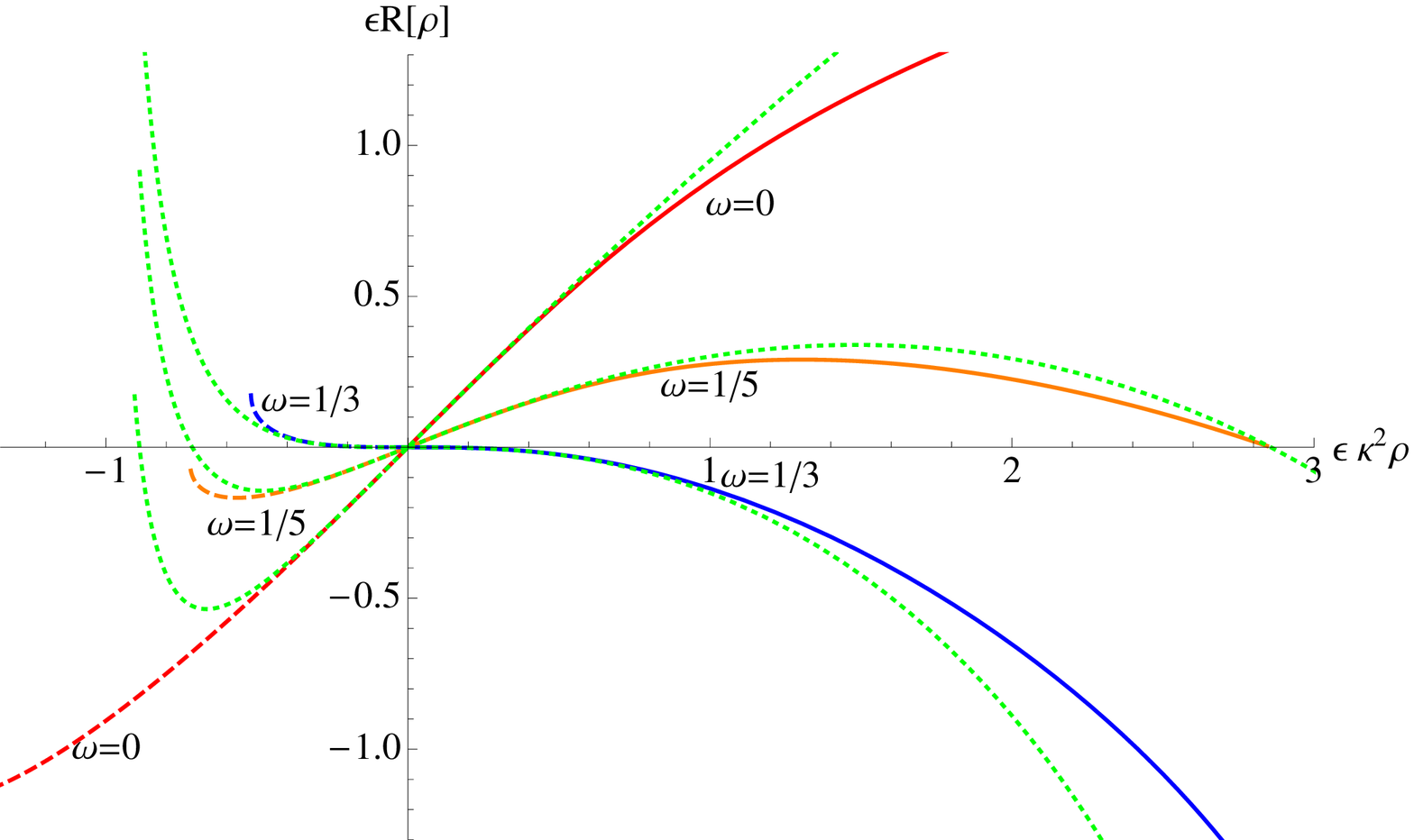}
\caption{Representation of the (dimensionless) Ricci scalar $\epsilon R$ as a function of the (dimensionless) energy density  $\epsilon \kappa^2\rho$ for the BI-$f(R)$ theory with $\alpha f(R)= -a\epsilon R^2/4$ and $a=1$. The dashed curves represent the bouncing solutions of the branch $\epsilon<0$. The solid curves correspond to the branch $\epsilon>0$.   The equations of state represent dust (red), radiation (blue), and a fluid with $\omega=1/5$ (orange). The green dotted lines represent the corresponding solutions in the original BI theory. 
 \label{Fig:Curv}}
\end{center}
\end{figure}

\section{Summary and conclusions \label{sec:summary}}

In this work we have considered a gravity theory formulated within the Palatini formalism consisting on a Born-Infeld-like gravitational Lagrangian plus an $f(R)$ term. This form of the gravity Lagrangian provides more flexibility to the original Born-Infeld theory, which possesses very interesting properties in scenarios involving cosmic as well as black hole singularities,  and allows to explore modifications of its dynamics at high  and low energies. We have provided a formal solution for the connection equation and a compact representation of the metric field equations. An algorithm that facilitates the analysis of perfect fluid cosmologies has also been worked out in detail and has been used to study some aspects of the high-energy dynamics of a specific model.  Our interest has focused on an $f(R)$ term of the form $f(R)\propto R^2$ which allows to tune at will the coefficient multiplying the $R^2$ term that arises in the low-energy series expansion of the Born-Infeld theory. This type of quadratic corrections are expected to arise due to the quantum properties of the matter fields in curved backgrounds. Depending on the number and types of matter fields \cite{Parker-Toms,Anderson}, the coefficient of the $R^2$ term may change, which justifies our study of this particular term.  The methods developed in this work are not restricted to the $R^2$ term and can also be applied to other $f(R)$ Lagrangians. \\

\begin{figure}[h]
\begin{center}
\includegraphics[width=0.75\textwidth]{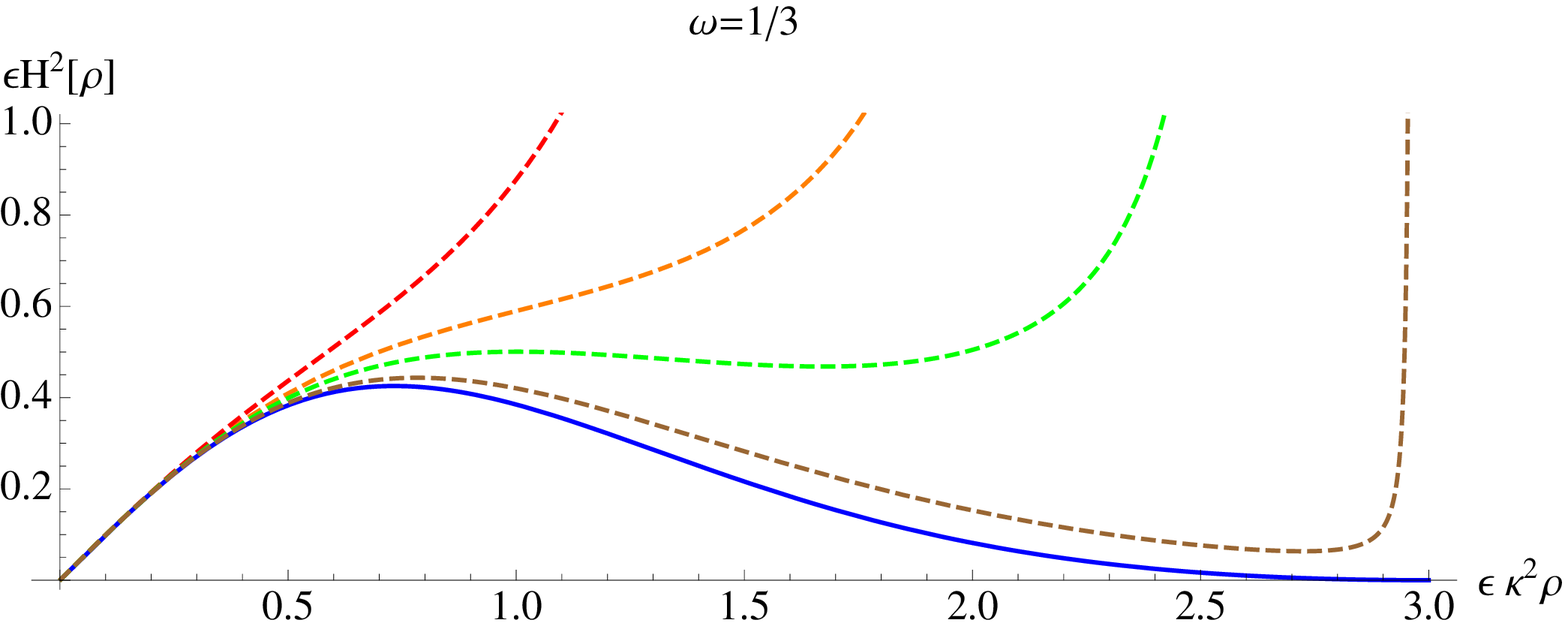}
\caption{Representation of the (dimensionless) Hubble function $\epsilon H^2$ as a function of the (dimensionless) energy density $\epsilon \kappa^2\rho$ for a radiation universe ($\omega=1/3$) in the cases $a=0$ (solid blue), $a=1/10$ (dashed brown), $a=1/3$ (dashed green), $a=1/2$ (dashed orange), and $a=1$ (dashed red). Note the long plateau following the local maximum around  $\epsilon \kappa^2\rho\approx 0.6$ in the case $a=1/3$, which could support a period of inflation generated by the radiation fluid.  \label{Fig:Inflation}}
\end{center}
\end{figure}

We have found that the solutions  with $\epsilon<0$, which yield a cosmic bounce, are robust against modifications of the $R^2$ coefficient, whereas those with $\epsilon>0$ undergo significant changes as compared to the original Born-Infeld theory. For equations of state $\omega>0$, the $\epsilon>0$ branch of Born-Infeld theory yields cosmologies with a stationary point characterized by $H^2=0$ and $dH^2/d\rho=0$. These solutions do not represent a bounce, but a state of minimum volume and maximum density that evolves into a standard FRW cosmology at late times. From Fig.\ref{Fig:Inflation} we see that any modification of the $R^2$ term in a radiation universe destroys the regularity of the original solution. However, the modifications experienced by these solutions may lead to a period of inflationary (de Sitter-like) expansion shortly after the big bang singularity, as is evident from the plateau of the curve $a=1/3$ in Fig.\ref{Fig:Inflation} and of the lower left curve with $a=1/2$ in Fig. \ref{Fig:Positive_Branch}. These results put forward that with slight modifications of the Born-Infeld theory one may get the conditions for an inflationary stage without the need for new dynamical degrees of freedom. Additional effects could be obtained by including higher-order powers of $R$ with free coefficients without altering the number of dynamical degrees of freedom of the theory. \\

The possibility of combining the Born-Infeld Lagrangian with an $f(R)$ term also offers new avenues to address a number of relevant questions of the gravitational dynamics at lower energies. In particular, one may look for $f(R)$ terms designed to modify the high-energy dynamics which combined with the Born-Infeld Lagrangian could leave a low-energy remnant in the form of an effective cosmological constant able to justify the late-time cosmic accelerated expansion. Another application could be the identification of $f(R)$ terms able to yield fully satisfactory models of stellar structure without the need to reconsider the convenient perfect fluid approximation \cite{Kim:2013nna,Olmo:2013gqa, Avelino:2012qe}, a currently open question that has attracted much attention from different perspectives. These and other questions will be considered elsewhere. \\

\section*{Acknowledgments}

GJO is supported by the Spanish grant FIS2011-29813-C02-02, the Consolider Program CPANPHY-1205388, the JAE-doc program and  i-LINK0780 grant of the Spanish Research Council (CSIC),  and by CNPq (Brazilian agency) through project No. 301137/2014-5. 
S.D.O. and A.N.M. are  supported by
the grant of Russian Ministry of Education and Science, project TSPU-139 and
the grant for LRSS, project No 88.2012.2.

\appendix
\section{Another example of the conformal approach. \label{App0}}

In this Appendix we illustrate the conformal approach in a different family of theories in which the departure from the BI theory is introduced via an $F(R)$ term but in a way that differs from that considered in this work so far. This new theory is defined by the following action
\be
\label{act2}
S_{{EiBI2}}=\frac{2}{\kappa}\int d^4x\left[\sqrt{|\det{\left(g_{\mu\nu}+\kappa R_{\mu\nu}(\Gamma)+\alpha g_{\mu\nu} F(R)\right)}|}-\lambda\sqrt{|g|}\right]
+S_M[g,\Psi] \ ,
\ee
where the notation is the same as in the restu of the paper. 
The connection equation for the  action (\ref{act2}) takes the form
\be
\label{equa2}
\nabla_\alpha\left[\sqrt{p}\left(\kappa\left( p^{-1}\right)^{\mu\nu}+\alpha   \left( p^{-1}\right)^{\sigma\rho} g_{\sigma\rho}  F'(R)g^{\mu\nu}\right)\right]=0.
\ee
Here $p_{\mu\nu}=g_{\mu\nu}+\kappa R_{\mu\nu}(\Gamma)+\alpha g_{\mu\nu} F(g^{\sigma\rho}R_{\sigma\rho}(\Gamma))$.
Variation of the metric yields
\be
\label{e2_1}
\sqrt{p}\left(p^{-1}\right)^{\mu\nu}(1+\alpha F(R))-
\alpha\sqrt{p}\left(p^{-1}\right)^{\sigma\rho} g_{\sigma\rho} F(R)' R^{\mu\nu}
-\lambda \sqrt{g}g^{\mu\nu}=-\kappa \sqrt{g}T^{\mu\nu}.
\ee
Imposing a conformal ansatz, 
\be
\label{p1}
p_{\mu\nu}=f(t) g_{\mu\nu} \ ,
\ee
we find that the auxiliary metric $u_{\mu\nu}$ that defines the connection 
\be
\label{metric2}
\Gamma^\alpha_{\mu\nu}=\frac{1}{2} u^{\alpha\beta}\left(\partial_\mu u_{\nu\beta}+\partial_\nu u_{\mu\beta}-\partial_\beta u_{\mu\nu}\right).
\ee
takes the form
\be
\label{up}
u_{\mu\nu}=f(t)(\kappa+\alpha n F'_R(R))  g_{\mu\nu}.
\ee

One can write the relationship between the scalar curvature and metric

\be
\label{Rup}
R_{\mu\nu}=\frac{1}{\kappa}\left[f-1-\alpha F(g^{\sigma\tau} R(u_{\sigma\tau}))\right]  g_{\mu\nu}.
\ee

Suppose that for the spatially-flat FRW universe with metric
\be
\label{FRW}
ds^{2}=-dt^{2}+a^{2}(t)(dx^{2}+dy^{2}+dz^{2})\, ,
\ee
the auxiliary metrics  takes the following form
\be \label{gg}
u_{\mu\nu}=u(t)\, diag(-1,a(t)^2,a(t)^2,a(t)^2).
\ee
Here  $u(t)=f(t)(\kappa+\alpha n F'_R(R))$. Suppose now that 
$R_{\mu\nu}=r(t) g_{\mu\nu}$ (the explicit form  $r(t)$ is easy to find from the expression (\ref{Rup})). Construct for the metric (\ref{gg}) the Christoffel symbols and Ricci tensor. Performing a calculation analogous to that for the original form of the action,we get another form of the function $F(R)$
\be
\label{ff2}
F(R)= -\frac{4+\kappa R\pm\sqrt{16\lambda+c R^2}}{4} \ ,
\ee
from which we obtain 
\be
p_{\mu\nu}=\mp\frac{\sqrt{\lambda n^2+c R^2}}{n} g_{\mu\nu}.
\ee
The action (\ref{act2}) takes then the form
\be \sqrt{|g_{\mu\nu} \frac{\sqrt{\lambda n^2+c R^2}}{n}|},
\ee
or, equivalently, 
\be \sqrt{|g_{\mu\nu}|} \frac{\lambda n^2+c R^2}{n^2}.
\ee
We thus find that in this case, the action (\ref{act2}) becomes
\be
S_{{EiBI2}}=\frac{2}{\kappa}\int d^4x\left[\sqrt{|g|}R^2\right].
\ee

\section{Non-conformal ansatz in vacuum. \label{App1}}
Let us assume now, in analogy with  (\ref{eq:con-var0}), that there exists a tensor $u_{\mu\nu}$ such that
$\nabla_\alpha(\sqrt{|u|}u^{\mu\nu})=0$. The connection equation for this theory then becomes
\be
\label{2.2}
\sqrt{|u|}\left(u^{-1}\right)^{\mu\nu}=\sqrt{|q|}\left( q^{-1}\right)^{\mu\nu}+\sqrt{g} g^{\mu\nu} f_R.
\ee
(in this section we set $\alpha=1$), which together with (\ref{eq:g-var0}) conforms the required system of equations. Assume now non-conformal ansatzes of the form $u_{\mu\nu}=$diag$(-u_0(t)^2,u_1(t)^2,u_1(t)^2,u_1(t)^2)$,  $q_{\mu\nu}=$diag$(-q_0(t)^2,q_1(t)^2,q_1(t)^2,q_1(t)^2)$ and that 
$g_{\mu\nu}$ has  a standard FLRW form. In this case, the tensors $u$ and $q$  can be expressed through the scalar curvature, and the function $f(R)$. We can get two different types of solutions of these equations. 
The first type is
\be
q_0=\pm\frac{\sqrt{-2 \lambda+\epsilon f(R)+2 f_R}}{\sqrt{-2+2 f_R}},q_1=\pm\frac{a \sqrt{-2 \lambda+\epsilon f(R)+2f_R}}{ \sqrt{-2+ 2f_R}}
\ee
for which $q_{\mu\nu}\sim g_{\mu\nu}$. This case was discussed above.

In the second case, the tensor $q_{\mu\nu}$ has the form
\be
q_0=\pm\frac{\sqrt{- \lambda+\frac{1}{2} \epsilon f(R)+f_R}}{\sqrt{f_R+{f_R}^3}},\,\,q_1=\mp \frac{a f_R \sqrt{- \lambda+\frac{1}{2} \epsilon f(R)+f_R}}{\sqrt{f_R+{f_R}^3}}
\ee
And connectivity between  tensors  $q_{\mu\mu}$  and $g_{\mu\nu}$ becomes more complex. For this case, one finds the following equation for the function $f(R)$:
\be
R+\frac{-f(R) \left(\epsilon+3 \epsilon {f_R}^2\right)+2 \left(\lambda+3{f_R}+3 \lambda{f_R}^2+{f_R}^3\right)}{2 \epsilon \left({f_R}+{f_R}^3\right)}=0 \ ,
\ee
which can be solved as
\begin{eqnarray}
f_1&=&
\frac{2 \lambda}{\kappa}
\pm\frac{\sqrt{6}}{9 \kappa}\sqrt{-\frac{1+\epsilon (R-2 c)+\sqrt{(1+\epsilon R)^2-4 \epsilon (\epsilon R-2) c+16 \epsilon^2 c^2}}{\epsilon c}} \times\nonumber\\
&\times&\left(2+2 \epsilon R-4 \epsilon c-\sqrt{(1+\epsilon R)^2-4 \epsilon (-2+\epsilon R) c+16 \epsilon^2 c^2}\right),
\end{eqnarray}
and
\begin{eqnarray}
f_2&=&
\frac{2 \lambda}{\epsilon}
\pm\frac{\sqrt{6}}{9 \epsilon}\sqrt{\frac{-1-\epsilon (R-2 c)+\sqrt{(1+\epsilon R)^2-4 \epsilon (\epsilon R-2) c+16 \epsilon^2 c^2}}{\epsilon c}} \times\nonumber\\
&\times&\left(2+2 \epsilon R-4 \epsilon c+\sqrt{(1+\epsilon R)^2-4 \epsilon (-2+\epsilon R) c+16 \epsilon^2 c^2}\right),
\end{eqnarray}
One can consider the following limit $R\to 0$, then we get
\be
f_1 \to \frac{2 \lambda}{\epsilon}\pm 2\frac{1-8\epsilon \,c}{9\epsilon}\sqrt{-3-\frac{3}{c\epsilon}},
\ee
\be
f_2 \to \frac{2( \lambda\pm 1)}{\epsilon}.
\ee

On the other hand, if $R\to \infty$, then
\be
f_1 \to \pm R^{3/2}\frac{2}{3\sqrt{-3c}},
\ee
\be
f_2 \to \pm 2 R^{1/2}\sqrt{c+\frac{1}{\epsilon}}.
\ee

\end{document}